\magnification1200

\rightline{KCL-MTH-12-08}

\vskip 2cm
\centerline
{\bf   $E_{11}$, generalised space-time and equations of motion in four
dimensions}
\vskip 1cm
\centerline{ Peter West}
\centerline{Department of Mathematics}
\centerline{King's College, London WC2R 2LS, UK}
\vskip 2cm
\leftline{\sl Abstract}
We construct the non-linear realisation of the semi-direct product of
$E_{11}$ and its first fundamental representation  at low
levels in four  dimensions. We include the fields for gravity, the
scalars and the gauge fields as well as the duals of these
fields. The generalised space-time, upon which the fields depend, 
consists of the usual coordinates of four dimensional space-time and
Lorentz scalar coordinates which belong to the 56-dimensional
representation of $E_7$. We demand that the equations of motion are
first order in derivatives of the generalised space-time and then show
that they are essentially uniquely determined by  the properties of the
 $E_{11}$ Kac-Moody algebra and its first fundamental representation.
The two lowest equations correctly describe the equations of motion
of the scalars and the gauge fields once one takes the fields to 
 depend only on the usual four dimensional space-time.

\vskip2cm
\noindent

\vskip .5cm

\vfill
\eject

\medskip
 {\bf {1 Introduction}}
\medskip
One of the most remarkable discoveries in the development of
supersymmetry, and indeed string theory,  was the presence of an $E_7$
symmetry in the four dimensional maximal
supergravity theory [1,2]. There followed the discovery of  $E_8$ [3] and
$E_9$ [4] symmetries in the maximal supergravity theories in three and two
dimensions respectively,  as well as a conjectured $E_{10}$ symmetry in
one dimension [5].  It was also found that the ten dimensional IIB
supergravity theory  possessed a SL(2) symmetry [6].  Apart from
the last symmetry it was universally assumed that these symmetries were a
quirk of dimensional reduction on a torus. The one exception was discussed
in the papers of references [7, 8]. The first of these  papers sacrificed
the  tangent space group to be just SO(1,3) but was then able to show that
eleven dimensional supergravity  possessed a SU(8) symmetry; the latter
papers in this reference were variations on this theme. While in 
reference [8]  it was argued that eleven dimensional supergravity  
possessed exceptional structures such as a generalised vielbein
associated with the group
$E_8$, however, the presence of these structures did not lead to the
conclusion that the theory possessed an $E_8$ symmetry group. 
\par
Inspired by the observation that the eleven dimensional supergravity
theory was a non-linear realisation [9] it was conjectured that the 
non-linear realisation of $E_{11}$ contained  eleven
dimensional supergravity [10]. This work was  extended to show that the
non-linear realisation of $E_{11}$ also contained the IIA [10], IIB [11]
and lower dimensional supergravity theories [12,13,14,15].  The
different theories result  from the different possible
decompositions of the Kac-Moody algebra 
$E_{11}$ that one could take. As the
maximal supergravity theories contain all   effects at low energy of the
underlying theory of strings and branes it was proposed  that this
underlying theory of strings and branes should possess an
$E_{11}$ symmetry [10]. In the early papers
our usual notion of space-time was introduced by  adjoining the space-time
translation generators  to the $E_{11}$ algebra in an adhoc step. 
However, in 2003 the non-linear realisation of  the semi-direct
product of $E_{11}$  together  with generators that belong to  its first
fundamental representation, denoted $l_1$, was considered [16]; this
algebra was denoted by
$E_{11}\otimes_s l_1$.  The highest weight state in the $l_1$
representations corresponds to the usual space-time translations, but
this representation contains an infinite number of elements. We recall
that the notion of a semi-direct product is well known to physicists as
the Poincar\'e group is just the semi-direct product of the Lorentz group
and the space-time translations.
 To understand reference [16], that is, the non-linear realisation of
$E_{11}\otimes_s l_1$ one has to be familiar with the notion
of a non-linear realisation which in this case is quite distinct from
what is often called a sigma model. This subject was once well known, at
least in some sections of the community and  particularly   in 
Russia  in the 1960's,  but this knowledge seems to have largely been
lost in the present,  with some notable exceptions. Examples of such
non-linear realisations can be found in [9,10,19] and  a review of
non-linear realisations, and the $E_{11}$ programme,  can be found in the
book of reference [17]. The non-linear realisation of the type considered
in [16]   introduces a generalised space-time which is automatically
equipped with a   generalised vielbein and corresponding
generalised tangent space. 
\par
The $l_1$ representation contains an infinite number of elements and so
introduces an infinite number of generalised coordinates. Like the
adjoint representations of $E_{11}$,  the elements of the $l_1$
representation can be organised according to the notion of a level [18].
The decomposition which leads  to  the $d$ dimensional theory, 
 is found by deleting node $d$ in the $E_{11}$ Dynkin diagram and
 at the lowest level    the resulting algebra is $GL(d)\times E_{11-d}$.
We recognise the
$GL(d)$ algebra as that associated with gravity in $d$ dimensions [19,10]
and $E_{11-d}$ as the U duality group in $d$ dimensions. How one finds 
the
$d$ dimensional theory is discussed in detail, for example,  in [12-15].
The generalised coordinates that arise from the $l_1$ representation at
the lowest level are the usual coordinates of  space-time as well as 
coordinates which are scalars under the Lorentz group but transform as  
the 10,
$\bar {16}$, 
$\bar {27}$, 56 and
$248\oplus 1$ of SL(5), SO(5,5),
$E_6$. $E_7$ and $E_8$ for $d$ equal to seven, six, five,  four and
three dimensions respectively  [20,21,13,22]. 
In fact one can   find all the generalised coordinates that are forms,
that is, carry completely anti-symmetrised space-time indices. This 
result for the generators of the $l_1$ representation, appropriate to 
$d$ dimensions, are given in the table below [20,21,22]

\bigskip

 {\centerline{\bf {The form charges in the $l_1$ representation  in d
dimensions}}}
\medskip
$$\halign{\centerline{#} \cr
\vbox{\offinterlineskip
\halign{\strut \vrule \quad \hfil # \hfil\quad &\vrule Ê\quad \hfil #
\hfil\quad &\vrule \hfil # \hfil
&\vrule \hfil # \hfil Ê&\vrule \hfil # \hfil &\vrule \hfil # \hfil &
\vrule \hfil # \hfil &\vrule \hfil # \hfil &\vrule \hfil # \hfil &
\vrule \hfil # \hfil &\vrule#
\cr
\noalign{\hrule}
D&G&$Z$&$Z^{a}$&$Z^{a_1a_2}$&$Z^{a_1\ldots a_{3}}$&$Z^{a_1\ldots a_
{4}}$&$Z^{a_1\ldots a_{5}}$&$Z^{a_1\ldots a_6}$&$Z^{a_1\ldots a_7}$&\cr
\noalign{\hrule}
8&$SL(3)\otimes SL(2)$&$\bf (3,2)$&$\bf (\bar 3,1)$&$\bf (1,2)$&$\bf
(3,1)$&$\bf (\bar 3,2)$&$\bf (1,3)$&$\bf (3,2)$&$\bf (6,1)$&\cr
&&&&&&&$\bf (8,1)$&$\bf (6,2)$&$\bf (18,1)$&\cr Ê&&&&&&&$\bf (1,1)$&&$
\bf
(3,1)$&\cr Ê&&&&&&&&&$\bf (6,1)$&\cr
&&&&&&&&&$\bf (3,3)$&\cr
\noalign{\hrule}
7&$SL(5)$&$\bf 10$&$\bf\bar 5$&$\bf 5$&$\bf \bar {10}$&$\bf 24$&$\bf
40$&$\bf 70$&-&\cr Ê&&&&&&$\bf 1$&$\bf 15$&$\bf 50$&-&\cr
&&&&&&&$\bf 10$&$\bf 45$&-&\cr
&&&&&&&&$\bf 5$&-&\cr
\noalign{\hrule}
6&$SO(5,5)$&$\bf \bar {16}$&$\bf 10$&$\bf 16$&$\bf 45$&$\bf \bar
{144}$&$\bf 320$&-&-&\cr &&&&&$\bf 1$&$\bf 16$&$\bf 126$&-&-&\cr
&&&&&&&$\bf 120$&-&-&\cr
\noalign{\hrule}
5&$E_6$&$\bf\bar { 27}$&$\bf 27$&$\bf 78$&$\bf \bar {351}$&$\bf
1728$&-&-&-&\cr Ê&&&&$\bf 1$&$\bf \bar {27}$&$\bf 351$&-&-&-&\cr
&&&&&&$\bf 27$&-&-&-&\cr
\noalign{\hrule}
4&$E_7$&$\bf 56$&$\bf 133$&$\bf 912$&$\bf 8645$&-&-&-&-&\cr
&&&$\bf 1$&$\bf 56$&$\bf 1539$&-&-&-&-&\cr
&&&&&$\bf 133$&-&-&-&-&\cr
&&&&&$\bf 1$&-&-&-&-&\cr
\noalign{\hrule}
3&$E_8$&$\bf 248$&$\bf 3875$&$\bf 147250$&-&-&-&-&-&\cr
&&$\bf1$&$\bf248$&$\bf 30380$&-&-&-&-&-&\cr
&&&$\bf 1$&$\bf 3875$&-&-&-&-&-&\cr
&&&&$\bf 248$&-&-&-&-&-&\cr
&&&&$\bf 1$&-&-&-&-&-&\cr
\noalign{\hrule}
}}\cr}$$
\par

The corresponding coordinates can be easily read off and  they carry the
contragredient representations to that of the generators. One sees  in
the first column the scalar coordinates mentioned above. The  generalised 
tangent space structure that is inherited from the  coordinates is easily
read off from the table in an obvious way. 
\par 
One can view the generalised coordinates from a more physical
viewpoint. The
$l_1$ representation can be thought of as containing all the brane charges
and so there is a one to one correspondence between the coordinates of the
generalised space-time and the brane charges [16,18,23,22,20]. As such one
can think of each coordinate as associated with a given type of brane
probe.  Furthermore for every
generator  in the Borel
subalgebra  of $E_{11}$ there is a corresponding element in the $l_1$
representation [18],  and  as a result   for every field in
the non-linear realisation one finds a corresponding coordinate. For
example,  in eleven dimensions at lowest level one has the usual field  of
gravity
$h_a{}^b$ associated with which one has the usual coordinate $x^a$ of 
space-time,  at the  level one we find the three form field
$A_{a_1a_2a_3}$
 with associated  coordinate $x_{a_1a_2}$, at level two we have the
six form field 
$A_{a_1\ldots a_6}$  with the  associated coordinate $x_{a_1\ldots a_5}$,
at level three the dual field of gravity $h_{a_1\ldots a_8, b}$ with a
corresponding  coordinate $x_{a_1\ldots a_7, b}$ and similarly at higher
levels [16]. One can think of this as generalisation of the notion of
space-time introduced by Einstein that takes into account the presence
of  fields, required by supersymmetry,  in addition to the metric. The
precise  correspondence,  for the four dimensional theory,  between the
fields and the coordinates can be found later in this paper. 
\par
Although quite a number of the predictions of $E_{11}$ have
been verified, see for example [13,17] for an account, the
radically new nature of the generalised space-time has, until relatively
recently, discouraged the systematic calculation of the $E_{11}\otimes_s
l_1$ non-linear realisation.
In the early papers  on
$E_{11}$ only the coordinate $x^a$ was used and the  symmetries of the
non-linear realisation were only implemented at lowest levels. This
particularly, applies to the local subalgebra which plays an important
part in the non-linear realisation. The local subalgebra is taken
to be the   Cartan involution invariant
subalgera and it was
usually taken to be just that at the lowest level which, in eleven
dimensions,  is   just the Lorentz algebra. As a result much of the power
of the non-linear realisation was lost. Nonetheless many of the features
of the supergravity theories were recovered. With retrospect one can view
the early attempts to construct the dynamics, see for example [10,11], as
using  the
$E_{11}\otimes_s l_1$ non-linear realisation but only keeping  the lowest
level $l_1$ generators, which,  in eleven dimensions,  are just the usual
space-time translations and so only the usual coordinates of space-time.  
\par
One of the first papers to use some of the higher level generalised
coordinates is given in reference [13] which 
was used to construct all gauged supergravities in five dimensions [13],
a result not previously known.  In this construction some of the
generalised coordinates and their corresponding components of the 
generalised vielbein played an important role.  However,  the
remaining coordinates of the $l_1$ representations were discarded and the
generalised  space-time that remained  was a 
slice taken in the $l_1$ representations  and $E_{11}$.  The
techniques used in this paper could easily be applied to find all gauged
supergravities in  all dimensions. 
\par
The construction of the $E_{11}\otimes_s l_1$ non-linear
realisation  at lowest level in four dimensions was carried out in
reference [24, 25]. This contained the usual coordinates of space-time
and Lorentz scalar coordinates which belonged to the 56-dimensional
representation of
$E_7$, mentioned above, that is the content of the $l_1$
representations at lowest level. It also took the fields in the $l_1$
representation at lowest level that is  the metric and the scalar fields. 
Much of these papers were devoted to the part of the theory that lives on
the 56-dimensional space.   In contrast to the
equation of motion approach pursued in most  $E_{11}$ papers, this paper
constructed  an invariant  Lagrangian. This Lagrangian was not uniquely
determined by the symmetries of the non-linear realisation,  which were
taken to be those at lowest level,  and  it contained several
undetermined constants.  However, it was realised in the papers of
reference [24,25] that if one restricted the dependence of the fields to
be only on the usual coordinates of space-time then, for a suitable choice
of the constants, the action was gauge and general coordinate invariant. 
More recently,  the
$E_{11}\otimes_s l_1$ non-linear realisations,  at lowest level, in the
$l_1$ representation and in  $E_{11}$, and also discarding 
 the fields and coordinates of the usual space-time, 
were  constructed in dimensions four to seven  and 
corresponding  Lagrangians were constructed [26].  These Lagrangians in
six and seven dimensions had previously been constructed [27] using the
coordinates introduced  into  the first quantised dynamics with the aim
of encoding duality symmetries [28].  However, it was apparent  [26] that
these were just the result of the non-linear realisation of 
$E_{11}\otimes_s l_1$ at lowest level which was then  further truncated in
the  way just mentioned. 
\par
One of the first papers to compute the $E_{11}\otimes_s l_1$ non-linear
realisation at higher levels and also keeping some of the higher level
symmetries was contained in references [29,30]  which computed the ten
dimensional IIA theory. It kept the $E_{11}$ fields at levels zero and
one which contained the fields of the NS-NS  and R-R
sectors of the IIA string respectively, with  the coordinates of the
$l_1$ representation at level zero. The latter consisted of the usual
coordinates of ten dimensional space-time as well as coordinates which
were all forms of odd rank.  Considering quantities that were  first order
in the  derivatives  of the generalised space-time it was shown that
there existed  only two covariant objects,  both of which were uniquely
determined and transformed into themselves. Setting one of these to zero
resulted in a set of equations of motion which were those of  type IIA
supergravity when the fields were taken to depend only on the usual
coordinates of ten dimensional space-time. The way the results 
in these papers was phrased were a
bit different but it is equivalent to the statement just made. 
\par
Very recently the non-linear realisation of $E_{11}\otimes_s l_1$ in
eleven dimensions was constructed  keeping the fields of gravity, three
form,  six form and dual gravity fields and the usual coordinates of
space-time as well as  the two form and five form coordinates [31]. In
other words,   this calculation kept the coordinates of the  $l_1$
representation and
$E_{11}$ up to and including levels two  and three respectively.
As a result one could impose  the higher level symmetries contained
in the $E_{11}\otimes_s l_1$ non-linear realisation and, in particular, 
the local symmetries. As in ten dimensions we considered quantities that
were first order in the derivatives of the generalised space-time and
again found that there existed only two such covariant objects both of
which were unique and transformed into themselves. Setting one of them to
zero resulted in a set of equations of motion,  the first of which 
correctly described the  equation of  motion
for the three form and six form fields when the dependence on the higher
level coordinates  and fields beyond the dual graviton were discarded. 
 The second equation in this set described the equation of motion
relating the usual field of gravity and the dual gravity field. This
equation was very close to being correct and one source of the
discrepancy  may be accounted for by missing contributions from higher
order fields which were omitted. We note that should one succeed in
finding an equation that correctly describes gravity then the
$E_{11}$ conjecture would be confirmed, that is, the non-linear
realisation of $E_{11}\otimes_s l_1$ is an extension of the maximal
supergravity theories. In this event one would have to take seriously the
generalised space-time associated with the
$l_1$ representation.  
\par
In this paper  we will carry out a similar calculation but in four
dimensions. That is we will carry out the $E_{11}\otimes_s l_1$
non-linear realisation in the decomposition that leads to the 
four-dimensional theory. We will keep the fields in the $l_1$
representation up to level four, that is we include the fields of
gravity, scalars,  one forms, two forms and the dual field of gravity.
From the  $l_1$ representation we will take the usual coordinates of
space-time and the Lorentz scalar coordinates that transform in the
56-dimensional representations of $E_7$, that is we only keep the level
zero part of the
$l_1$ representation. However, we will do the calculation in such a way
that we will impose some of  the crucial higher level symmetries of the
non-linear realisation. The $E_{11}\otimes_s l_1$ algebra for the eleven
dimensional theory is relatively simple and well studied up to the
required levels however, the complicated index structure of the fields
means that the calculation of the equations of motion and the
verification of their invariance, given in [31],  is   rather intricate.
Although the four dimensional theory has more fields compared to the
eleven dimensional theory they have a much simple index structure. Hence
although it is much more complicated to find the $E_{11}\otimes_s l_1$
algebra for the required decomposition,  the calculation of the equations
of motion is much simpler. To find the four dimensional theory
one must  decompose $E_{11}$ and the $l_1$ representations into
representations of 
$GL(4)\otimes E_7$.  While this is relatively straightforward, in order to
correctly implement even the lowest order local subalgebra of the
non-linear realisation  one must then further decompose into
representations of  
$SO(1,3)\otimes SU(8)$. This complicated calculation  takes up
much of this paper, however, there are many checks one can carry out to
verify that the results found are correct. One further advantage of 
four dimensions is that the usual gravity field and the dual gravity
field, and  their related coordinates,  have similar index structures. 
As such one is hopefully in a better position to resolve the
problems associated with the dual gravity field.
\par
We will consider how objects that are first order in derivatives with
respect to the generalised space-time transform under the symmetries of
the non-linear realisation. We will show that there are two sets of
objects that transform into themselves and are uniquely specified by the
symmetries. Each set contains an infinite number of objects and setting
one set to zero leads at the lowest level to the correct equations of
motion for the scalars and gauge fields provided we take the fields to
only depend on the  coordinates of the usual four dimensional space-time. 
We will also give some preliminary results on the higher level fields. 
\par
None of the  papers mentioned above gives a satisfactory account of 
how the familiar space-time we are used to
emerges naturally from the generalised space-time, or put another way, 
how does one discard most of the coordinates of the $l_1$ representation
to find the theories we are familiar with. This dilemma is considered 
further in the discussion section. 
\par
Before concluding the introduction we will make a few remarks on the
relation of the $E_{11}$ programme to other approaches. We begin by  
 contrasting the usual view of  M theory with the
$E_{11}$ conjecture. It is often stated that all the different string
theories are different aspects of an eleven dimensional theory,  called M
theory. Of course we do not know much about M theory so the meaning of
this statement is not  clear. In contrast all the maximal
supergravity theories can be obtained from the $E_{11}\otimes_s l_1$
non-linear realisation by taking different decompositions. As such
from the $E_{11}$ viewpoint all the different theories are on an equal
footing and indeed are dual to each other,  being  different descriptions 
of the same underlying theory,  and indeed no space-time dimension is
preferred. The mapping between the different theories is for example
discussed in reference [23]. 
\par
Subsequent to the 2003 proposal of reference [16] a number of other
approaches involving some kind of  generalised geometry have been
considered.   The most popular is called doubled geometry, see [32] and
references therein. This approach was inspired by earlier papers that
considered in addition to the usual coordinate of space-time $x^a$ the
coordinate
$y_a$  and constructed an O(10,10), T duality,  invariant  ten
dimensional theory. This theory had the same fields as the NS-NS
sector of the superstring but they depended on both of the just
mentioned coordinates. However, the result was none other than the
$E_{11}\otimes_s l_1$ non-linear realisation suitable for the IIA theory
at lowest level [29].  The advantage of viewing it this way is that it is
part of a much large conceptual framework where all the symmetries are
automatically encoded. Indeed  the
non-linear realisation of
$E_{11}\otimes_s l_1$ at level one leads to the inclusion of the R-R
sector
 fields [30].  The level zero and one calculations of the non-linear
realisation appropriate for the IIA theory are very simple and can be
performed in a few pages without the need for any guess work.
The one point contained in the literature on doubled field theory which
does not follow from  the $E_{11}\otimes_s l_1$ non-linear realisation is
how to discard the $y_a$ coordinates. Rather than just discard them as in 
references [25,26] they adopt what is called a section condition, 
however, in practice it seems there is not so much difference. 
\par
There is yet another approach  inspired by the work of references 
 [33] and  [34]. This introduced an extended tangent
space, associated with O(D,D) but does  not extend our usual notion of
space-time, see for example  [35] and references therein. In this
approach one does not try to find additional symmetries, but rather
 packages the theory up into a generalised geometry. In ten dimensions
the tangent space is essentially doubled compared to that taken
conventionally, however this is precisely the same tangent space as
arises in the   non-linear realisation
of
$E_{11}\otimes_s l_1$ appropriate to ten dimensions at lowest level 
[29,30].  While in lower
dimensions the tangent spaces and tangent groups are just those contained
in the $E_{11}\otimes_s l_1$ non-linear realisation at lowest level
[20,21,13,22].  While there has
not been a detailed  study to investigate the connection to  the
non-linear realisation of $E_{11}\otimes_s l_1$ it would seem inevitable
that  it is just the non-linear realisation of $E_{11}\otimes_s l_1$
with the
$l_1$ part taken to be just the usual coordinates of space-time and the
$E_{11}$ part up to  the level  that incorporates all the usual
supergravity fields.  Thus, while  the works of reference [35] try to
generalise Einstein geometry taking into account global considerations, 
the required structures are very likely to be  automatically
encoded in the 
$E_{11}\otimes_s l_1$  non-linear realisation.

\medskip
 {\bf {2 The $E_{11}$ algebra and $l_1$
representation viewed from  four dimensions}}
\medskip

In this section we will formulate the $E_{11}$ algebra and the $l_1$
representation in such a way that their  non-linear realisation
 leads to a theory in four dimensions. We will begin by 
finding their decompositions in terms of representations of $GL(4)\otimes
E_{7}$ rather than the more common  SL(11) decomposition which leads to
the eleven dimensional theory. 
An important role in the construction of the non-linear realisation, that
is the dynamics, is played by the Cartan involution invariant
subalgebra of $E_{11}$, denoted $I_c(E_{11})$. At lowest
level this is just the subalgebra $SO(1,3)\otimes SU(8)$. The $SU(8)$
factor is just the Cartan involution invariant subgroup of $E_7$  while
$SO(1,3)$ is the Cartan involution invariant subalgebra of SL(4); these
are the same as the respective  maximal compact subgroup for the  real
forms of $E_7$ and $A_3$  with which we are working.   As such we need a
formulation of $E_{11}$ and the $l_1$ representation in which the
subalgebra
$I_c(E_{11})$ is apparent and in particular the subalgebra 
$SO(1,3)\otimes SU(8)$. While  it is obvious how
representations of SL(4) can be rewritten in terms of SO(1,3) this is not
quite so clear for the SU(8) hidden within $E_7$. This problem has been
well studied in the mathematics literature and an account for physicists
can be found in appendix B of reference [2]. 
As explained in this reference the best way to find this SU(8) subalgebra 
of $E_7$ is to first identify the more obvious SL(8) subgroup of $E_7$ and
then decompose the adjoint representations of $E_7$ in terms
of representations of this SL(8). The desired  SU(8) subalgebra,  and
the decomposition of the adjoint representation of $E_7$ into
representations of it, can then be constructed. We note that  the SL(8)
and SU(8) have  in common their  obvious SO(8) subgroups, but  they are
not different real forms of the same subalgebra of $E_7$ when viewed in
its complex form. 
\par
For the calculations in this paper we must generalise  this 
results to find,   at low levels,  the Cartan involution
subalgebra of $E_{11}$, that is $I_c(E_{11})$,  and then decompose
$E_{11}$, and its $l_1$ representations, into representations of
$I_c(E_{11})$. This is the task of section two. At lowest level this  was
carried out in reference [26], that is, the SU(8) contained within
$E_{11}$ was identified and the $E_7$ and 56-dimensional representation
in $l_1$  were decomposed in terms of this SU(8). 
\medskip
 {\bf {2.1 The $GL(4)\otimes E_{7}$ decomposition  of the $E_{11}$ algebra
and the $l_1$ representation}}
\medskip
To find the four-dimensional theory from the non-linear realisation of
$E_{11}\otimes_s l_1$ we delete the fourth node in the $E_{11}$ Dynkin
diagram, see figure 2.1,  and consider the decomposition of
$E_{11}\otimes_s l_1$ into representations of the  subalgebra that
results, that is, $GL(4)\otimes E_{7}$. 
$$
\matrix{
&&&&&&& & & & &&& &\bullet &11&&&\cr 
&&&&&& & & & &&& & &| & && & \cr
\bullet &-&\bullet &-&\bullet&-&\otimes&-&\bullet&-&\bullet  &-
&\bullet&-&\bullet&-&\bullet&-&\bullet
\cr
1& &2&&3 &&4 &&5&&6& &7& &8& & 9&
&10\cr}
$$
\medskip

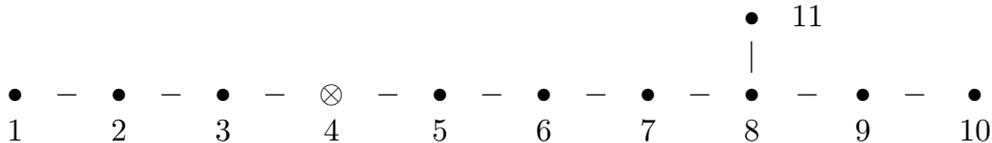
Figure 2.1: The $E_{11}$ Dynkin diagram from the viewpoint of 
 the four-dimensional theory.
\medskip

The GL(4) factor, whose generators we denote by $K^a{}_b, \
a,b=1,\ldots, 4$ leads in the non-linear realisation to the familiar 
field used to describe gravity and the
$E_{7}$, whose generator we denote by $R^\alpha$,   is the well known
symmetry group in four dimensions whose corresponding fields in the
non-linear realisation are the seventy scalars. 
\par
The representations that occur in this decomposition can be classified
according to a level. In this case  the level is just the number of upper
minus lower GL(4) indices that the generator possess. Thus the level zero 
$E_{11}$ generators are just $GL(4)\otimes E_{7}$. The interested reader
can find a more formal account of the level in earlier $E_{11}$
papers and in the book [17].  Ordered by their level the 
decomposition of
$E_{11}$ into representations of $GL(4)\otimes E_{7}$ is given by 
$$
K^a{}_b (15, 1,0) ,\  R^\alpha (1, 133, 0);\  R^{aN} (4, 56, 1) ;\ 
R^{a_1a_2\alpha} (6, 133, 2),$$
$$
\  \hat K ^{ab} (10, 1, 2) ;\ 
R^{a_1a_2a_3\lambda} (4, 912, 3), \  R^{a_1a_2, b N} (20, 56, 3); \ldots 
\eqno(2.1.1)$$
where $\ldots$ indicate generators at level four and above. 
The  first two figures in the brackets
indicate the dimensions of the SL(4) and
$E_7$ representations respectively, while the last  figure is the level.
We have  not displayed the negative level generators,  except for those
at level zero, however, they have the same index structure except that
their indices are now subscripts. 
The GL(4) indices are given as $a,b,  a_1,
a_2,\ldots $. The indices on the  generators $R^{a_1a_2\alpha}$ and
$R^{a_1a_2a_3\lambda} $ are totally anti-symmetrised. The generator $
\hat K ^{ab}$  is subject to the condition $\hat K ^{ab} =\hat K
^{(ab)}$ and the generator  $R^{a_1a_2, b N}$ satisfies  $  R^{[ a_1a_2,
b] N}=0$. 
\par
As explained later, in the non-linear realisation there is a one to one
correspondence between the fields that arise and the generators of the
Borel subalgebra, that is those of level zero and positive level. The
$GL(4)\otimes
E_{7}$ indices on the fields are inherited from those on these generators,
for example  $R^{aN}A_{aN}$. As such, the generators with completely
anti-symmetrised indices, the so called form generators, lead to the gauge
fields $A_{aN}, A_{a_1a_2\alpha}, A_{a_1a_2a_3\lambda}$.  One expects, in
the resulting dynamics,  that the first gauge fields will satisfy some
kind of  self-duality condition, the second gauge fields are dual to the
scalars and the third lead to a cosmological constant and so classify  the
gauged supergravities [12,36].  As we have mentioned the usual field
$h_a{}^b$ of gravity corresponds to the generator $K^a{}_b$ while the
generator
$\hat K^{ab}$ leads to the dual gravity field, denoted by  $\hat h_{ab}$.
In this paper we will be only concerned with the generators up to, and
including,  level two. 
\par
We also decompose the $l_1$ representation into representations of
$GL(4)\otimes E_{7}$. Listed according to their level the results is 
$$
P_a (4,1,0), Z^N (1, 56, 1), Z^{a \alpha} (4, 133, 2), 
Z^a (4,1,2), Z^{a_1 a_2 N} (6,56,3),
$$
$$ Z^{(a_1 a_2) N} (10,56,3),
Z^{a_1 a_2\lambda} (6,912, 3) ,\dots 
\ldots 
\eqno(2.1.2)$$
where $\ldots $ denoted objects at level four and above. In the non-linear
realisation these lead to the generalised space-time  which  has 
the corresponding coordinates 
$$
x^a, x_N, x_{a\alpha},\hat x_{a}, x_{a_1a_2 N}, x_{(a_1 a_2)N} ,
x_{a_1a_2\lambda},\ldots 
\ldots 
\eqno(2.1.3)$$
We note that  for the four dimensional theory, the gravity and dual
gravity fields and usual coordinates $x^a$ and dual gravity coordinates
$\hat x_a$ appear on a much more  symmetrical footing than in other
dimensions. 
\par 
One can find the  generators of equation (2.1.1) and (2.1.2) by simply
"dimensionally reducing"  to four dimensions the
$E_{11}$ algebra, and the $l_1$ representation when written  in their 
eleven dimensional formulations. From the group
theoretic view point this is equivalent to decomposing the SL(11)
formulation of the $E_{11}$  algebra and the $l_1$ representation into
representations of
$GL(4)\otimes SL(7)$. The SL(11) in question is the algebra that
result from deleting node eleven in the $E_{11}$  Dynkin diagram of
figure 2.1 and so its Dynkin diagram  consists of nodes one to ten. We
recall that the $GL(4)\otimes E_7$ subalgebra arose from deleting node
four in the Dynkin diagram of figure 2.1 and  the 
 the  SL(4) and SL(7)  subalgebras correspond to the   Dynkin diagrams
which consist of nodes one to three and nodes five to ten respectively.
Having found the decomposition into $GL(4)\otimes SL(7)$ representations
we can then repackage the SL(7) representations into those of $E_7$.  The
eleven dimensional, or  SL(11), formulation of  the
$E_{11}$ algebra is well known, and when listed according to the
appropriate level, it contains  the positive level generators [10,37]
$$
K^{\hat a}{}_{\hat b}\ (0), R^{\hat a_1\hat a_2\hat a_3}\ (1),  
R^{\hat a_1\ldots \hat a_6}\ (2), R^{\hat a_1\ldots\hat a_8,\hat b}\ (3),
$$
$$ 
R^{\hat a_1\ldots\hat a_9,\hat b_1\hat b_2\hat b_3}\ (4), 
R^{\hat a_1\ldots\hat a_{11},\hat b}\ (4), R^{\hat a_1\ldots\hat
a_{10},(\hat b_1\hat b_2 )}\ (4), 
\ldots 
\eqno(2.1.4)$$
where $\hat a, \hat b=1,\ldots , 11$. In this equation $\ldots
$ denotes  generators at level five and higher. The generators obey
irreducibility conditions such as 
$R^{[\hat a_1\ldots\hat a_8,\hat b]}=0, \ldots $. The numbers in the
brackets indicate the level appropriate to the SL(11) decomposition. 
 This level arises from deleting node eleven and it is different to the
level discussed above that is associated with  deleting node four. 
The eleven dimensional level can be thought of as  just the number of
up minus down eleven dimensional indices divided by three. 
\par
The
$l_1$ representation,  listed according to increasing level,  contains
[16,18,20] 
$$
P_{\hat a} \ (0),\  Z^{\hat a_1\hat a_2}\ (1),
\  Z^{\hat a_1\ldots \hat a_5}\ (2)
,\  Z^{\hat a_1\ldots \hat a_7,b}\ (3),
\  Z^{\hat a_1\ldots \hat a_8}\ (3),
$$
$$
\  Z^{\hat b_1\hat b_2 \hat b_3, \hat a_1\ldots \hat a_8}\ (4),
\  Z^{(\hat c \hat d ), \hat a_1\ldots \hat a_9}\ (4),
\  Z^{\hat c\hat d,\hat a_1\ldots \hat a_9}\ (4),\ 
\  Z^{\hat c,\hat a_1\ldots \hat a_{10}}\ (4),\ 
Z\ (4)
$$
$$
Z^{\hat c, \hat d_1\ldots \hat d_4,\hat a_1\ldots \hat a_9}\ (5),\ 
Z^{\hat c_1\ldots \hat c_6,\hat a_1\ldots \hat a_8}\ (5),\ 
Z^{\hat c_1\ldots \hat c_5,\hat a_1\ldots \hat a_9}\ (5),\ 
$$
$$
Z^{\hat d_1,\hat c_1 \hat c_2 \hat c_3,\hat a_1\ldots \hat a_{10}}\
(5),\  Z^{\hat c_1 \ldots \hat c_4,\hat a_1\ldots \hat a_{10}}\
(5,-2),\  Z^{(\hat c_1\hat c_2,\hat c_3 )}\ (5),\ 
Z^{\hat c,\hat a_1\hat a_2}\  (5),\ 
\ldots 
\eqno(2.1.5)$$
These generators satisfy irreducibility conditions  such as $Z^{[\hat
a_1\ldots\hat a_7,\hat  b]}=0, \ldots $. 
\par
To carry out the decomposition from representations of SL(11) into those
of $GL(4)\otimes SL(7)$ we 
  divide the values of the indices $\hat a, \hat b, \ldots$ into the
ranges,  one to four and the remainder, that is,  five to eleven,
which we denote by the labels $a$ and $i$ respectively. Carrying this out
for the
$l_1$ representation given in equation (2.1.5) we find that 
$$
P_a; \dot P_i, \dot Z^{i_1i_2}, \dot Z^{i_1\ldots i_5} , \dot Z^{i_1\ldots
i_7, j};\dot  Z^{a i},Z^{a i_1\ldots i_4},\dots 
\eqno(2.1.6)$$
We have placed a dot on  the resulting objects as we will use the
symbols $ Z^{i_1i_2}$ etc for a later purpose. We now have to assemble
these into representations of
$GL(4)\otimes E_7$. Those of GL(4) are the same, but  we must
assemble the representations of SL(7) into those of  $E_7$.  For
example,  between the first  semi-colon and the second,  we
recognise the $56=7+21+21+7$-dimensional  representation of $E_7$
contained in the 
$Z^N$ of equation (2.1.2). 
\par
Proceeding in a similar way we can decompose the generators of the
$E_{11}$ algebra of equation (2.1.4) into representations of SL(7) to
find 
$$
K^a{}_b; \dot K^i{}_j, \dot  R^{i_1i_2i_3},\dot  R_{i_1i_2i_3}, \dot
R^{i_1\ldots i_6},\dot  R_{i_1\ldots i_6}; \dot K^a{}_i,  \dot R^{a
i_1i_2},
\dot R^{a i_1\ldots i_5}, \dot R^{ i_1\ldots i_7, a}: \dots 
\eqno(2.1.7)$$ 
The first generators before the first semicolon are those of the 
$15+1$-dimensional representation of GL(4). The generators between
the next semicolons are those of the
adjoint representations of $E_7$ corresponding to the decomposition 
$133= 48+1+35+35+7+7$ and
those after the next semicolon belong to the
$56=7+21+21+7$-dimensional representation of
$E_7$; thus we find  agreement  with equation (2.1.1). 
\medskip
 {\bf {2.2 The SL(8) Formulation for the $E_{11}$ algebra and $l_1$
representation}}
\medskip
As we have explained at the beginning of  section two an  important role
in the construction of the non-linear realisation 
is played by the Cartan involution invariant subalgebra of $E_{11}$ which
at lowest level  is  the subalgebra $SO(1,3)\otimes SU(8)$. To
locate this algebra we first identify the SL(8) subgroup of $E_7$ whose
generators we denote by $K^I{}_J, I,J=1,\ldots , 8$. The obvious  SL(7)
subgroup of SL(8) is  the SL(7) subgroup discussed in the
last section. In terms of the  generators of equation (2.1.7) the
generators of SL(8) are given by [26]
$$
K^i{}_j= \tilde K^i{}_j -{1\over 6} \delta _j^i \sum_k \tilde K^k{}_k ,
\quad i,j=1,\ldots ,7
$$
$$K^{8}{}_{j}=-{2\over6!}\epsilon_{ji_{1}\cdots i_{6}}\dot R^{i_{1}\cdots
i_{6}}, 
\quad 
K^{j}{}_{8}={2\over6!}\epsilon^{ji_{1}\cdots i_{6}}\dot R_{i_{1}\cdots
i_{6}}.
\eqno(2.2.1)$$
where 
$$
\tilde K^i{}_j= \dot  K^i{}_j- {1\over 2}\delta ^i_j \sum_a K^a{}_a
\eqno(2.2.2)$$
On the right hand side of these  equations  the symbols $\dot
K^i{}_j,
\dot R^{i_{1}\cdots i_{6}}$ and $\dot R_{i_{1}\cdots
i_{6}}$ are those found by the decomposition given in equation (2.1.7) and
should not be confused with the $K^i{}_j$ on the left hand side of
equation (2.2.1) which are part of the SL(8) generators. 
It is straightforward to verify, using the $E_{11}$ algebra given in
appendix A,  that they do indeed satisfy the SL(8) algebra, that is, 
$$
[K^{I}{}_{J},K^{L}{}_{M}]=\delta^{L}_{J}K^{I}{}_{M}-\delta^{I}_{M}K^{L}{}_{J}
\eqno(2.2.3)$$
Since we are dealing with SL(8) the generators are traceless and so  
$K^8{}_8=-\sum_{i=I}^7 K^I{}_I$
\par
The remaining generators of $E_7$ belong to the seventy-dimensional
representation of SL(8) and are contained in the 
generator $R^{I_1\ldots I_4}$ whose indices are totally antisymmetric. In
terms of the decomposed generators of equation (2.1.7) these are given by 
$$
R^{i_{1}i_{2}i_{3}8}={1\over12 }\dot R^{i_{1}i_{2}i_{3}}
, \quad 
R^{i_{1}\ldots i_{4}}={1\over 12. 3!}\epsilon^{i_{1}\cdots
i_{4}j_{1}j_{2}j_{3}}\dot R_{j_{1}j_{2}j_{3}}
\eqno(2.2.4)$$
The commutators of these generators with those of 
SL(8) are given by 
$$
[K^{I}{}_{J},R^{L_{1}\cdots
L_{4}}]=4\delta^{[L_{1}}_{J}R^{|I|L_{2}\cdots
L_{4}]}-{1\over2}\delta^{I}_{J}R^{L_{1}\cdots L_{4}}
\eqno(2.2.5)$$
while the remaining $E_7$ commutators are given by 
$$
[R^{I_{1}\cdots I_{4}},R^{J_{1}\cdots
J_{4}}]=-{1\over8}\{\epsilon^{J_{1}\cdots J_{4}L[I_{1}\cdots
I_{3}}K^{I_{4}]}{}_{L}-\epsilon^{I_{1}\cdots I_{4}L[J_{1}\cdots
J_{3}}K^{J_{4}]}{}_{L}\}
$$
$$=-{1\over4}\epsilon^{J_{1}\cdots J_{4}L[I_{1}\cdots
I_{3}}K^{I_{4}]}{}_{L}.
\eqno(2.2.6)$$
In deriving the last relation we have used the identity 
$$
\epsilon^{J_{1}\cdots J_{4}L[I_{1}\cdots
I_{3}}S^{I_{4}]}{}_{L}+
\epsilon^{I_{1}\cdots I_{4}L[J_{1}\cdots
J_{3}}S^{J_{4}]}{}_{L}=-\epsilon^{J_{1}\cdots J_{4}I_{1}\cdots
I_4}\sum_N S^N{}_{N}
\eqno(2.2.7)$$
valid for any object $S^I{}_{J}$. This identity is easily proved by
taking values for the indices. 
\par
Proceeding in a similar way one finds that the positive level generators
of the 
$E_{11}$ algebra, when written in terms of
$GL(4)\otimes SL(8)$ representations,  take the form 
$$
 K^I{}_J (1, 63), R^{I_1\ldots I_4} (1, 70), K^a{}_b (16, 1);
 R^{a I_1I_2} (4, 28), R^{a}{}_{I_1 I_2} (4, \bar { 28}) ;
$$
$$
R^{a_1a_2 I}{}_J (6, 63), R^{a_1a_2 I_1\ldots I_4} (6, 70), \hat K^{ab}
(10, 1),\ldots 
 \eqno(2.2.8)$$
The two numbers in brackets give the dimensions of their SL(4) and SL(8) 
representations respectively. The negative definite level 
generators
  are given by  
$$
\tilde  R_{a I_1I_2} (4,\bar { 28}),\tilde  R_{a}{}^{I_1 I_2} (4,  28) ;
\tilde R_{a_1a_2}{}^ I{}_J (6, 63),\tilde  R_{a_1a_2}{}^{ I_1\ldots I_4}
(6, 70),
\tilde {\hat K}_{ab} (10, 1),\ldots  
 \eqno(2.2.9)$$
\par 
 The level one
generators of equation (2.2.8) are identified with the underlying $E_{11}$
algebra of equation (2.1.7)  as follows
$$
R^{a}{}^{ i_1i_2}=\dot R^{a i_1i_2},\  R^{a j8}= {2\over  7!}\epsilon
_{i_1\ldots i_7} \dot R^{a i_1\ldots i_7, j}
$$
$$
R^{a}{}_{ i_1i_2}= -{2\over 5!}\epsilon
_{i_1i_2j_1\ldots i_5} \dot R^{a j_1\ldots j_5},\  R^a{}_{i8}= -\dot
K^a{}_i
\eqno(2.2.10)$$
while for the level minus one generators the identification is given by 
$$
\tilde R_{a i_1i_2}=\dot R_{a i_1i_2},\  \tilde R_{a j8}= {2\over 
7!}\epsilon ^{i_1\ldots i_7} \dot R_{a i_1\ldots i_7, j}
$$
$$
\tilde R^{a i_1i_2}= -{2\over 5!}\epsilon
^{i_1i_2j_1\ldots i_5} \dot R_{a j_1\ldots j_5},\ \tilde  R_a{}^{i8}= \dot
K^i{}_a
\eqno(2.2.11)$$
\par
 Using the above  identifications and the $E_{11}$ commutators
given in appendix A  one can deduce that the commutators of the $E_{11}$
algebra when written in terms of the $GL(4)\otimes SL(8)$
decomposition, that is,  as  given in  equations (2.2.8) and
(2.2.9). The commutators  with the SL(8) generators 
of
$E_7$ are given by 
$$
[K^I{}_J, R^{a K_1 K_2}] = 2\delta _J^{[ K_1 |}R^{a I |K_2]}
-{1\over 4} \delta^I_J  R^{a K_1 K_2}
\quad 
[K^I{}_J, R^{a}{}_{ K_1 K_2}] =- 2\delta ^I_{[ K_1 |}R^{a}{}_{ J |K_2]}
+{1\over 4} \delta^I_J  R^{a}{}_{ K_1 K_2}
\eqno(2.2.12)$$
together with analogous result for the other generators. The commutators
with the $R^{I_1\ldots I_4}$ generators of $E_7$ are 
$$
[R^{I_1\ldots I_4}, R^{a J_1 J_2}] = {1\over 4!} \epsilon^{I_1\ldots
I_4J_1J_2 K_1K_2}R^{a}{}_{K_1K_2}
\quad  
[R^{I_1\ldots I_4}, R^{a}{}_{ J_1 J_2}] = \delta ^{[I_1 I_2}_{J_1 J_2}
R^{a}{}^{ I_3I_4]}
$$
$$
[R^{I_1\ldots I_4}, R^{a_1a_2 J}{}_K ]= -4\delta ^{[I_1|}_K
R^{a_1a_2 J| I_2I_3I_4]} +{1\over 2} \delta ^J_K R^{a_1a_2I_1\ldots
I_4},
$$
$$
[R^{I_1\ldots I_4},\tilde  R_{a_1a_2}{}^{ J}{}_K ]= 4\delta _{[I_1|}^K
\tilde R_{a_1a_2}{}_{ J| I_2I_3I_4]} -{1\over 2} \delta ^J_K
\tilde R_{a_1a_2}{}_{I_1\ldots I_4},
$$
$$
[R^{I_1\ldots I_4}, R^{a_1a_2 J_1\ldots J_4 ]}]={1\over 36} \epsilon
^{I_1\ldots I_4 L [ J_1J_2J_3|} R^{a_1a_2|J_4 ]}{}_L
=- {1\over 36} \epsilon^{J_1\ldots J_4 L [ I_1I_2I_3|} R^{a_1a_2|I_4
]}{}_L
$$
$$
[R^{I_1\ldots I_4}, \tilde R_{a_1a_2}{}_{ J_1\ldots J_4 ]}]=-{2\over 3} 
\delta ^{[ I_1I_2I_3}_{  [ J_1J_2J_3} \tilde R_{a_1a_2}{}^{I_4 ]}{}_{J_4]}
\eqno(2.2.13)$$
\par
The commutators of the level one generators with themselves are given by 
$$ 
[R^{aI_{1}I_{2}},R^{bI_{3}I_{4}}]=-12 R^{abI_{1}\cdots I_{4}}, \quad 
[R^{aI_{1}I_{2}},R^{b}{}_{J_{1}J_{2}}]
=+4\delta^{[I_{1}}_{[J_{1}}R^{|ab|I_{2}}{}_{J_{2}]}
+2\delta^{I_{1}I_{2}}_{J_{1}J_{2}}\hat K^{ab}
$$
$$
[R^{a}_{I_{1}I_{2}},R^{b}_{J_{1}J_{2}}]= {1\over 2}
\epsilon_{I_{1}I_{2}J_{1}J_{2}K_1\ldots K_4} R^{a_1a_2 K_1\ldots K_4}=
12\star  R^{a_1a_2}{}_{ I_{1}I_{2}J_{1}J_{2}}
\eqno(2.2.14)$$
where $\star R^{a_1a_2}{}_{ I_1\ldots I_4}= {1\over 4!}
\epsilon_{I_1\ldots I_4 J_1\ldots J_4} R^{a_1a_2 J_1\ldots J_4}$. 
While the equivalent 
commutators  for the level minus one generators with themselves are 
given by 
$$
[\tilde{R}_{aI_{1}I_{2}},\tilde{R}_{bJ_{1}J_{2}}]=-12\star
\tilde{R}_{abI_{1}I_{2}J_{1}J_{2}},
\quad
[\tilde{R}_{aI_{1}I_{2}},\tilde{R}_{b}{}^{J_{1}J_{2}}]
=+4\delta^{[J_{1}}_{[I_{1}}\tilde{R}_{ab}{}^{J_{2}]}{}_{I_{2}]}
+2\delta^{J_{1}J_{2}}_{I_{1}I_{2}}\tilde{\hat K}_{ab}$$
$$[\tilde{R}_{a}^{I_{1}I_{2}},\tilde{R}_{b}{}^{J_{1}J_{2}}]
={12}\tilde{R}_{ab}{}^{I_{1}I_{2}J_{1}J_{2}}
\eqno(2.2.15)$$
\par
The commutators between the level one and minus one
generators are given by 
$$
[R^{aI_{1}I_{2}},\tilde{R}_{bJ_{1}J_{2}}]
=2\delta^{I_{1}I_{2}}_{J_{1}J_{2}}K^{a}_{b}
+4\delta^{a}_{b}\delta^{[I_{1}}_{[J_{1}}K^{I_{2}]}{}_{J_{2}]}
-\delta^{a}_{b}\delta^{I_{1}I_{2}}_{J_{1}J_{2}}\sum^{4}_{c=1}K^{c}{}_{c}$$
$$[R^{aI_{1}I_{2}},\tilde{R}_{b}{}^{J_{1}J_{2}}]
=-12\delta^{a}_{b}R^{I_{1}I_{2}J_{1}J_{2}} , \quad
[R^{a}_{I_{1}I_{2}},\tilde{R}_{bJ_{1}J_{2}}]=12\delta^{a}_{b}\star
R_{I_{1}I_{2}J_{1}J_{2}}$$
$$[R^{a}{}_{I_{1}I_{2}},\tilde{R}_{b}{}^{J_{1}J_{2}}]
=-2\delta^{J_{1}J_{2}}_{I_{1}I_{2}}K^{a}{}_{b}
+4\delta^{a}_{b}\delta_{[I_{1}}^{[J_{1}}K^{J_{2}]}{}_{I_{2}]}
+\delta^{a}_{b}\delta_{I_{1}I_{2}}^{J_{1}J_{2}}\sum^{4}_{c=1}K^{c}{}_{c}
\eqno(2.2.16)$$ 
Finally we list the commutators between the level two and
minus one generators. 
$$
[R^{abI}{}_{J},\tilde{R}_{cK_1K_2}]
=-4\delta^{[a}_{c}\delta^{I}_{[K_1|}R^{b]}{}_{J|K_2]}+ {1\over 2} 
\delta^{[a |}_{c}\delta^{I}_J R^{| b]}{}_{K_1K_2},
$$
$$
[R^{abI}{}_{J},\tilde{R}_{c}{}^{K_1K_2}]
=4\delta^{[a}_{c}\delta^{[K_1 |}_{J} R^{b]I|K_2]}
- {1\over 2} \delta^{[a |}_{c}\delta^{I}_J R^{| b]}{}^{K_1K_2}$$
$$[R^{abI_{1}\cdots
I_{4}},\tilde{R}_{cK_{1}K_{2}}]
=2\delta^{[a|}_{c}\delta^{[I_{1}I_{2}}_{K_{1}K_{2}}R^{|b]I_{3}I_{4}]},
\quad
[R^{abI_{1}\cdots
I_{4}},\tilde{R}_{c}{}^{K_{1}K_{2}}]={1\over 12}\epsilon^{I_{1}\cdots
I_{4}K_{1}K_{2}J_{1}J_{2}}\delta^{[a}_{c}R^{b]}{}_{J_{1}J_{2}}
$$
$$
[\hat K^{ab},\tilde R_c{}_{J_1J_2}]=- \delta ^{[a|}
_{c}R^{|b]}{}_{J_1J_2},\quad 
[\hat K^{ab},\tilde R_c{}^{J_1J_2}]=- \delta ^{[a|}
_{c}R^{|b]}{}^{J_1J_2}
\eqno(2.2.17)$$
and the level minus two and one generators 
$$
[\tilde R_{ab}{}^I{}_{J},{R}^{cK_1K_2}]
=-4\delta^{[c}_{[a}\delta^{[K_1 |}_{[J|}\tilde R_{b]}{}^{I|K_2]}
+ {1\over 2} 
\delta_{[a|}^{c}\delta^{I}_J\tilde R_{ | b]}{}^{K_1K_2}
$$
$$
[\tilde R_{ab}{}^I{}_{J},{R}^{c}{}_{K_1K_2}]
=4\delta_{[a |}^{c}\delta_{[K_1 |}^{I}\tilde R_{b] |J| K_2]}
- {1\over 2} 
\delta_{[a|}^{c}\delta^{I}_J\tilde R_{ | b]}{}_{K_1K_2}
$$
$$[\tilde R_{ab}{}^{I_{1}\cdots
I_{4}},{R}^{c}{}_{K_{1}K_{2}}]
=2\delta_{[a|}^{c}\delta^{[I_{1}I_{2}}_{K_{1}K_{2}}\tilde
R_{|b]}{}^{I_{3}I_{4}]}$$
$$[\
\tilde R_{ab}{}^{I_{1}\cdots
I_{4}},{R}^{c}{}^{K_{1}K_{2}}]={2\over 4!}\epsilon^{I_{1}\cdots
I_{4}K_{1}K_{2}J_{1}J_{2}}\delta_{[a}^{c}\tilde R_{b]}{}_{J_{1}J_{2}}
$$
$$
[\tilde {\hat K}_{ab}, R^c{}_{J_1J_2}]=- \delta _{[a|}
^{c}\tilde R_{|b]}{}_{J_1J_2},\quad 
[\tilde {\hat K}_{ab}, R^c{}^{J_1J_2}]=- \delta _{[a|}
^{c}\tilde R_{|b]}{}^{J_1J_2}\eqno(2.2.18)$$
\par
The identification with the level two generators given in equation
(2.2.8)  in the eleven dimensional formulation of $E_{11}$ is given by 
$$
R^{a_1a_2 i}{}_{8}= \dot R^{a_1a_2 i},\quad 
R^{a_1a_2 i}{}_{j}={2\over 6!} \epsilon _{jl_1\ldots l_6 } \dot
R^{a_1a_2\l_1\ldots l_6, i}- {\delta ^i_j \over 2.6!} 
\epsilon _{l_1\ldots l_7 } \dot
R^{a_1a_2\l_1\ldots l_6, l_7}
$$
$$
R^{a_1a_2 8}{}_{j} = {d_1\over 7!} \epsilon_{l_1\ldots l_7} \dot
R^{a_1a_2 l_1\ldots l_7}{}_{, j} \ , \quad 
R^{a_1a_2 i_1\ldots i_4} = -{1\over 2.3} \dot R^{a_1a_2 i_1\ldots
i_4},
$$ 
$$
R^{a_1a_2 i_1i_2 i_3 8} = -{d_2\over 7!}\epsilon_{l_1\ldots l_7}  \dot
R^{a_1a_2 l_1\ldots l_7, i_1i_2 i_3},\quad 
\hat K^{ab}= {2\over 7} \epsilon_{l_1\ldots l_7} \dot R^{(a| l_1\ldots
l_7, |b)}
\eqno(2.2.19 )$$
The constants $d_1$ and $d_2$ are yet to be fixed but they are not needed
for the derivation of the above commutators. We note that $R^{a_1a_2
8}{}_8= -\sum _k R^{a_1a_2 k}{}_k$
\par
We now write the $l_1$ representation in terms of SL(8)
representations using similar techniques. In terms of these
representations equation (2.12) can be written as 
$$
P_a (4,1) ; Z^{IJ} (1,28), Z_{IJ}(1, \bar {28}), \ldots
\eqno(2.2.20)$$
The identification with the SL(7) representations of equation
(2.1.7) is given by 
$$
Z^{ij}=\dot Z^{ij},\quad Z^{j8}={1\over 3. 7!}\epsilon_{i_{1}\cdots
i_{7}}\dot Z^{i_{1}\cdots i_{7},j},\quad 
$$
$$Z_{i_1i_2}={1\over
5!}\epsilon_{i_{1}i_{2}j_{1}\cdots j_{5}}\dot Z^{j_{1}\cdots
j_{5}},\quad  Z_{i8}=P_{i}
\eqno(2.2.21)$$
Using these identifications and the commutators of appendix A we find the
commutation relations with the generators of $E_7$ are given by 
$$
[R^{I_{1}\cdots I_{4}},Z^{J_{1}J_{2}}]={1\over
4!}\epsilon^{I_{1}\cdots
I_{4}J_{1}J_{2}K_{1}K_{2}}Z_{K_{1}K_{2}},
\quad [R^{I_{1}\cdots
I_{4}},Z_{J_{1}J_{2}}]=\delta^{[I_{1}I_{2}}_{J_{1}J_{2}}Z^{I_{3}I_{4}]}
$$
$$[K^{I}{}_{J},Z^{L_{1}L_{2}}]=\delta^{L_{1}}_{J}Z^{IL_{2}}-\delta^{L_{2}}_{J}Z^{IL_{1}}-{1\over
4}\delta^{I}_{J}Z^{L_{1}L_{2}}
, \quad [K^{I}{}_{J},P_a]=0
$$
$$[K^{I}{}_{J},Z_{L_{1}L_{2}}]=-\delta_{L_{1}}^{I}Z_{JL_{2}}+\delta_{L_{2}}^{I}Z_{JL_{1}}+{1\over
4}\delta^{I}_{J}Z_{L_{1}L_{2}}
\eqno(2.2.22)$$ 
Their commutators with the level one $E_{11}$ generators are given by 
$$
[R^{a}{}^{I_1I_2},  P_b]= \delta _b^a Z^{I_1I_2}, \quad 
[R^{a}{}_{I_1I_2},  P_b]= \delta _b^a Z_{I_1I_2}
\eqno(2.2.23)$$  
and with those at level minus one by 
$$[\tilde R_{a}{}_{I_1I_2},  Z^{J_1J_2}]= 2\delta ^{I_1I_2}_{J_1J_2} P_a
, \quad 
[\tilde R_{a}{}_{I_1I_2},  Z_{J_1J_2}]= 0, \quad [\tilde
R_{a}{}_{I_1I_2}, P_a]=0
$$
$$
[\tilde R_{a}{}^{I_1I_2},  Z^{J_1J_2}]=0,\quad 
[\tilde R_{a}{}^{I_1I_2}, 
Z_{J_1J_2}]= -2\delta _{I_1I_2}^{J_1J_2} P_a 
, \quad [\tilde
R_{a}{}^{I_1I_2}, P_a]=0
\eqno(2.2.24)$$ 
\medskip
 {\bf {2.3 The Cartan involution invariant subalgebra of $E_{11}$ in
four dimensions}}
\medskip 
In the last section we found the $E_{11}$ algebra and the $l_1$
representations in terms of $GL(4)\otimes SL(8)$ representations. We now
use this result to find the Cartan involution invariant subalgebra of
$E_{11}$ denoted $I_c(E_{11})$. A discussion of the Cartan involution can
be found in earlier papers on $E_{11}$ or in the book [17]. 
 The first step is to find the Cartan involutions
invariant subalgebra of $E_7$.  In terms of the SL(8) formulation of the
$E_{11}$ algebra of equation (2.2.8), the Cartan involution,  denoted by
$I_c$ ($I_c^2=I$),   acts on the level zero generators as 
$$
I_c(K^I{}_J)=- K^J{}_I,\ I_c(R^{I_1\ldots I_4})=-\star R^{I_1\ldots
I_4}\equiv -{1\over 4!}\epsilon^{I_1\ldots I_4J_1\ldots J_4}R^{J_1\ldots
J_4}, 
\eqno(2.3.1)$$
 on the level one
generators as 
$$
I_{c}(R^{aI_{a}I_{2}})=-\tilde{R}_{aI_{1}I_{2}},\quad
I_{c}(R^{a}{}_{I_{1}I_{2}})=\tilde{R}_{a}{}^{I_{1}I_{2}}
\eqno(2.3.2)$$
and on the level two generators as 
$$
I_{c}(R^{a_{1}a_{2}I}{}_{J})=-\tilde{R}_{a_{1}a_{2}}{}^{J}{}_I,\quad
I_{c}(R^{a_{1}a_{2}I_{1}\cdots I_{4}})=\star
\tilde{R}_{a_{1}a_{2}}{}^{I_{1}\cdots I_{4}}, \quad
I_{c}(\hat K^{ab})=-\tilde{\hat K}_{ab}
\eqno(2.3.3)$$
\par
One way to find these results is to use the identifications with the
generators of the eleven dimensional theory, given in  the previous
section,  and the known action of the Cartan involution on the
 generators. In particular, we have that 
$$
I_c(K^{\hat a}{}_{\hat b})=-K^{\hat b}{}_{\hat a}, \quad 
I_c ( R^{\hat a_1\hat a_2\hat a_3})  = - R_{\hat a_1\hat a_2\hat
a_3}, 
$$
$$
 I_c(  R^{\hat a_1\ldots \hat a_6})=   R_{\hat a_1\ldots
\hat a_6},\quad I_c ( R^{\hat
a_1\ldots\hat a_8,\hat b})= - R_{\hat a_1\ldots\hat a_8,\hat b}
$$
The positive sign in the third equation may seem incongruous, but it 
depends how one defines the generator $R_{\hat a_1\ldots
\hat a_6}$ and the original papers made an unforutnate choice that we are
now stuck with. 
For example, using equation (2.2.11) we find
that 
$$
I_c(R^{ai_1i_2})= I_c(\dot R^{ai_1i_2})= -\dot R_{ai_1i_2}= 
-\tilde  R_{ai_1i_2}
\eqno(2.3.4)$$
or, using equation (2.2.4),  we find that 
$$
I_{c}(R^{i_{1}i_{2}i_{3}8})=I_{c}({1\over 12} \dot R^{i_{1}i_{2}i_{3}})=
-{1\over 12}\dot R_{i_{1}i_{2}i_{3}}=
-{1 \over
4!}\epsilon_{i_{1}i_{2}i_{3}j_{1}\cdots
j_{4}}R^{j_{1}j_{2}j_{3}j_{4}}=-\star R^{i_{1}i_{2}i_{3}8}
$$
\par
At level zero the Cartan involution invariant subalgebra contains
the generators 
$$
J^I{}_J= K^I{}_J-K^J{}_I,\quad {\rm and }\quad 
S^{I_1\ldots I_4}= R^{I_1\ldots I_4}-\star R^{I_1\ldots I_4}
\eqno(2.3.5)$$
as well as the four dimensional Lorentz generators 
$$
J^a{}_b=K^a{}_b-K^b{}_a
\eqno(2.3.6)$$
We note that $\star S^{I_1\ldots I_4}=-S^{I_1\ldots I_4}$. The generators
of equation (2.3.5) obey  the algebra 
$$
[S^{I_{1}\cdots I_{4}},S_{J_{1}\cdots J_{4}}]
=-6\delta^{[I_{1}\cdots
I_{3}}_{[J_{1}\cdots J_{3}}J^{I_{4}]}{}_{J_{4}]}+{1 \over
4}\epsilon^{I_{1}\cdots I_{4}L[J_{1}J_{2}J_{3}}J^{J_{4}]}{}_{L}
\eqno(2.3.7)$$
as well as the commutation relations with $J^I{}_J$, which generate
SO(8),  and act on other generators in the expected way, for example 
$$
[J^I{}_J, S^{K_1\ldots K_4}]= 4\delta _J^{[K_1}R^{|I|K_2K_3K_4 ]}- 
4\delta _I^{[K_1}R^{|J|K_2K_3K_4 ]} 
\eqno(2.3.8)$$
These commutators of equation (2.3.7) and (2.3.8) are those of SU(8).
This is to be expected as SU(8) is well known to be the Cartan involution
invariant subalgebra of $E_7$. 
\par
At level plus and minus one the invariant generators 
are given by 
$$
S^{aI_1I_2}= R^{aI_1I_2}- \tilde R_{aI_1I_2}, \quad {\rm and }\quad 
\hat S^{a}{}_{I_1I_2}= R^{a}{}_{I_1I_2}+ \tilde R_{a}{}^{I_1I_2}
\eqno(2.3.9)$$
These 56 generators must transforms as the $28+\bar {28}$-dimensional 
representations of SU(8). However, the above generators mix under the
SU(8) commutators and the combinations that transform independently, that
is, irreducibly,  are given by 
$$
S^{aI_1I_2}_{\pm}= S^{aI_1I_2}\pm i \hat S^{a}{}_{I_1I_2}
\eqno(2.3.10)$$
Indeed we find that 
$$
[S^{I_{1}\cdots I_{4}},S_{\pm}^{aJ_{1}J_{2}}]=\mp{i \over
4!}\epsilon^{I_{1}\cdots
I_{4}J_{1}J_{2}K_{1}K_{2}}S^a_{\pm}{}_{K_{1}K_{2}}\pm i
\delta_{J_{1}J_{2}}^{[I_{1}I_{2}}S^a_{\pm}{}^{ I_{3}I_{4}]}
\eqno(2.3.11)$$
\par
The Cartan involution invariant generators at level plus and minus two
are 
$$
S^{a_{1}a_{2}I}{}_{J}= 
R^{a_{1}a_{2}I}{}_{J}-\tilde{R}_{a_{1}a_{2}}{}^{J}{}_I,\quad
S^{a_{1}a_{2}I_{1}\cdots I_{4}}= R^{a_{1}a_{2}I_{1}\cdots I_{4}}+\star
\tilde{R}_{a_{1}a_{2}}{}^{I_{1}\cdots I_{4}}, \quad
S^{ab}= \hat K^{ab}-\tilde{\hat K}_{ab}
\eqno(2.3.12)$$
We note that $S^{a_{1}a_{2}K}{}_{K}=0$, but otherwise this object has no
particular symmetry. Similarly, unlike $S^{I_{1}\cdots I_{4}}$, which is
anti-self dual $S^{a_{1}a_{2}I_{1}\cdots I_{4}}$ is neither self-dual or
anti-self dual.  Their commutators with the SO(8) generators is as one
expects and those with the remaining SU(8) generators are given by 
$$
[S^{I_1\ldots I_4}, S_S^{a_1a_2} {}^J{}_K]=Ê
-2\delta_K^{[I_1 |} S_+^{a_1a_2}{}^{J|I_2I_3I_4]}
-2\delta_J^{[I_1|} S_+^{a_1a_2}{}^{K|I_2I_3I_4]}
+{1\over 2} \delta_K^J S_+^{a_1a_2}{}^{I_1I_2I_3I_4]}
$$
$$
[S^{I_1\ldots I_4}, S_A^{a_1a_2} {}^J{}_K]=Ê
-2\delta_K^{[I_1|} S_-^{a_1a_2}{}^{J|I_2I_3I_4]}
+2\delta_J^{[I_1|}  S_-^{a_1a_2}{}^{K|I_2I_3I_4]}
$$
$$
[S^{I_1\ldots I_4}, S_+^{a_1a_2} {}^{J_1\ldots J_4}]=Ê
{4\over 3}  \delta^{[J_1J_2J_3 |}_{[I_1I_2I_3}
S_S^{a_1a_2}{}^{|J_4]}{}_{I_4]} +{1\over 18}  \epsilon ^{I_1\ldots I_4
L[J_1J_2J_3 |} S_S^{a_1a_2}{}^{|J_4]}{}_{L}
$$
$$
[S^{I_1\ldots I_4}, S_-^{a_1a_2} {}^{J_1\ldots J_4}]=Ê
{4\over 3}  \delta^{[J_1J_2J_3 |}_{[I_1I_2I_3}
S_A^{a_1a_2}{}^{|J_4]}{}_{I_4]} +{1\over 18}  \epsilon ^{I_1\ldots I_4
L[J_1J_2J_3 |} S_A^{a_1a_2}{}^{|J_4]}{}_{L}
\eqno(2.3.13)$$
where 
$$
S^{ab}_S{}^{I}{}_{J} = {1\over 2} (S^{ab}{}^{I}{}_{J}+
S^{ab}{}^{J}{}_{I}) , \quad 
S^{ab}_A{}^{I}{}_{J} = {1\over 2} (S^{ab}{}^{I}{}_{J}-
S^{ab}{}^{J}{}_{I}), 
$$
$$
S^{ab}_\pm {}^{I_1I_2I_3I_4}= S^{ab} {}^{I_1I_2I_3I_4}\pm \star
S^{ab} {}^{I_1I_2I_3I_4}
\eqno(2.3.14)$$ 
We can interpret these equations as meaning thatÊ
$S_+^{a_1a_2} {}^{J_1\ldots J_4}$ and  $S_S^{a_1a_2} {}^J{}_K$ form the
$35+35=70$-dimensional representations of SU(8) whileÊ
$S_-^{a_1a_2} {}^{J_1\ldots J_4}$ and  $ S_A^{a_1a_2} {}^J{}_K$ form
the $35+28=63$-dimensional representations of SU(8).ÊWe recall that these
generators belong to the 133-dimensional representation of $E_7$. In fact
the last term on the right-hand side of the last commutator in equation
(2.3.13) is zero as $L=J_4$ and the object is antisymmetric. 
\par
The commutators of the Cartan involution invariant generators of equation
(2.3.10) must give those of equations (2.3.5) and (2.3.12) and using the
commutators of section (2.2) we find that 
$$
[ S^{aI_1I_2}_{\pm}, S^b_\pm {}^{J_1J_2} ]= -12 S^{ab}_+ {}^
{I_1I_2J_1J_2}
\pm 8i \delta ^{[I_1}_{[J_1}
S^{ab}_S{}^{I_2]}{}_{J_2]} 
$$
$$
[ S^{aI_1I_2}_{\pm}, S^b_\mp {}^{J_1J_2} ]= -12 S^{ab}_-{}^{I_1I_2J_1J_2}
-8\delta_a^b\delta ^{[I_1}_{[J_1} J^{I_2]} {}_{J_2 ]}-4\delta ^{I_1I_2}
_{J_1J_2}J^a{}_b
$$
$$
\mp 8i \delta ^{[I_1} _{[J_1}
S^{ab}_A{}^{I_2]}{}_{J_2] }
\mp4i \delta ^{I_1I_2} _{J_1J_2}S^{ab}\pm 24i \delta _a^b
S^{I_1I_2J_1J_2}
\eqno(2.3.15)$$
While the commutators of the generators of equation (2.3.10) and those of
equation (2.3.12), but keeping only those generators of levels plus and
minus one,  are given by 
$$
[ S^{aI_1I_2}_{\pm}, S^{b_1b_2}_S{}^J{}_K]= \mp 4i \delta _a^{[b_1}
\delta _{[I_1}^{(K|} S^{|b_2]} _\mp {}_{|J) |I_2 ]}\pm {i\over 2} 
\delta ^J_K \delta _a^{[b_1}
 S^{b_2]} _\mp {}_{I_1 I_2 }, 
$$
$$
[ S^{aI_1I_2}_{\pm}, S^{b_1b_2}_A{}^J{}_K]= \pm 4i \delta _a^{[b_1}
\delta _{[I_1}^{[J|} S^{|b_2]} _\pm {}_{|K] |I_2 ]} , 
\eqno(2.3.16)$$ 
$$
[ S^{aI_1I_2}_{\pm}, S^{b_1b_2}_+{}^{J_1\ldots J_4} ]= 2\delta
^{[b_1}_a\{ \delta ^{[J_1J_2 |}_{I_1I_2} S^{	|b_2]}_\mp {}^{|J_3J_4]} +
{1¤\over 4!}
\epsilon^{J_1\ldots J_4 I_1I_2 K_1K_2} S^{| b_2]}_\mp {}_{K_1K_2}\}
$$
$$
[ S^{aI_1I_2}_{\pm}, T^{b_1b_2}_-{}^{J_1\ldots J_4} ]= 2\delta
^{[b_1}_a\{ \delta ^{[J_1J_2 |}_{I_1I_2} S^{	|b_2]}_\pm {}^{|J_3J_4]} -
{1\over 4!}
\epsilon^{J_1\ldots J_4 I_1I_2 K_1K_2} S^{| b_2]}_\pm {}_{K_1K_2}\}
$$
$$
[S^{aI_1I_2}_{\pm}, \hat T^{b_1b_2} ]= 
\pm i \delta_a^{[b_1} T^{b_2] I_1I_2}_{\pm}
\eqno(2.3.17)$$
The full commutators would contain in addition generators at level
plus and minus three which are beyond the level we are keeping. 
\par 
Finally we give the transformations of the
$l_1$ representation under the  Cartan involution invariant subalgebra.
Rather than work with
$Z^{I_1I_2}$ and $Z_{I_1I_2}$ we will work with the irreducible
representations of SU(8) which are given by 
$$
X_\pm{}^{I_1I_2}= Z^{I_1I_2}\pm i Z_{I_1I_2}
\eqno(2.3.18)$$
Using equation (2.2.22) we find that their  commutators with  the
generators of SU(8) are given by 
$$
[S^{I_1\ldots I_4}, X_\pm{}^{J_1J_2}]= \pm i  \delta _{[J_1J_2
|}^{[ I_1I_2}
 X_\pm{}^{I_3I_4]}]\mp {1\over 4!}\epsilon^{I_1\ldots I_4J_1J_2K_1K_2} 
X_\pm{}^{K_1K_2}
\eqno(2.3.19)$$
that is,  as the $28+\bar {28}$ of SU(8) should. 
\par
Using equations (2.2.23) and (2.2.24) we  also find that 
$$
[ S_\pm^a{}^{I_1I_2}, P_b]= \delta _b^a X_\pm{}^{I_1I_2},\quad 
[ S_\pm^a{}^{I_1I_2}, X_\pm^a{}^{J_1J_2}]=0,\quad
[ S_\pm^a{}^{I_1I_2}, X_\mp^a{}^{J_1J_2}]=-4  \delta _{J_1J_2}^{I_1I_2}P_a
\eqno(2.3.20)$$


\medskip
 {\bf {2.4 The decomposition of $E_{11}$ into  representations of its
Cartan invariant subalgebra }}
\medskip 
The generators of the $E_{11}$ algebra can be split into those that are
invariant under the Cartan involution and those that transform with a
minus sign. The former are those in the Cartan involution invariant
algebra $I_c(E_{11})$ given in the previous section. The latter are
sometimes called the coset generators and are the subject of this
section. Clearly, the commutator of two
elements of $I_c(E_{11})$ gives a result in $I_c(E_{11})$. Furthermore 
if $S\in I_c(E_{11})$ and $T$ is a coset generator, that is $I_c(T)=-T$, 
then their commutator
$[S, T]$ is also a coset generator since $I_c ([S, T])= [I_c(S), I_c(T)]=
-[S, T]$. As a result, the coset generators belong to a representations
of  $I_c(E_{11})$. In this section we find the commutators of
the coset generators with those of  $I_c(E_{11})$. Put another way we
wish to decompose the adjoint representation of $E_{11}$ into
representations of $I_c(E_{11})$.  In the next section we will use these 
commutation relations to deduce  the
crucial field variations under which the theory must be invariant.  
\par
Using equation (2.3.1) we find that the level zero the coset
generators  are given by 
$$ 
T^{I_{1}\cdots I_{4}}=R^{I_{1}\cdots I_{4}}+\star R^{I_{1}\cdots
I_{4}}, \quad
T^{I_{1}}{}_{I_{2}}=K^{I_{1}}{}_{I_{2}}+K^{I_{2}}{}_{I_{1}}, 
\eqno(2.4.1)$$
We note that $\sum_L T^L{}_L=0$ and  $\star T^{I_{1}\cdots
I_{4}}= T^{I_{1}\cdots I_{4}}$. Their commutators with the generators of
SO(8) are obvious and those with the remaining generators of SU(8) are 
given by 
$$
[S^{I_{1}\cdots I_{4}}, T^J{}_K ]= -4\{\delta _K^{[I_1} T^{|J|I_2\ldots
I_4]}-{1\over 4!}\epsilon_{I_{1}\cdots I_{4}KL_1L_2L_3}T^{JL_1L_2L_3}\}
+(K\leftrightarrow J)
$$
$$= \{-8 \delta _K^{[I_1} T^{|J|I_2\ldots
I_4]}+\delta _K^J T^{I_1I_2\ldots
I_4}\} +(K\leftrightarrow J)
\eqno(2.4.2)$$
and 
$$
[S^{I_{1}\cdots I_{4}}, T^{J_1\ldots J_4}]= 6\delta
_{[J_1J_2J_3}^{[I_1I_2I_3} T^{I_4]}{}_{J_4]}+
{1\over 4}\epsilon ^{I_1\ldots I_4L [J_1J_2J_3}T^{J_4]}{}_{L}
\eqno(2.4.3)$$
The generators of equation (2.4.1) belong to the 70-dimensional
representations of SU(8). 
\par
The coset generators formed from the level one and minus one generators
are given by 
$$
T^{a}{}^{I_{1}I_{2}}=R^{a}{}^{I_{1}I_{2}}+\tilde{R}_{a}{}_{I_{1}I_{2}},
\quad
\hat{T}^{a}_{I_{1}I_{2}}=R^{a}{}_{I_{1}I_{2}}-\tilde{R}_{a}{}^{I_{1}I_{2}}
\eqno(2.4.4)$$
As for the analogous objects in the Cartan involution invariant
subalgebra,  these two objects transform into each other under SU(8) and
so we define  instead the generators 
$$
T_{\pm}^{a}{}^{I_{1}I_{2}}=T^{a}{}^{I_{1}I_{2}}\pm i
\hat T^{a}{}_{I_{1}I_{2}}
\eqno(2.4.5)$$
Their commutators with the generators of SU(8) are given by 
$$
[S^{I_{1}\cdots I_{4}},T_{\pm}^{a}{}^{J_{1}J_{2}}]=\mp{i \over
4!}\epsilon^{I_{1}\cdots
I_{4}J_{1}J_{2}K_{1}K_{2}}T_{\pm}^a{}_{K_{1}K_{2}}\pm i
\delta^{J_{1}J_{2}}_{[I_{1}I_{2}}T_{\pm}^a{}_{ I_{3}I_{4}]}
\eqno(2.4.6)$$
and we recognise that they belong to the $28+\bar{28}$ representations of
SU(8). We note that the SU(8) representations do not emerge in the
familiar form, for example   the 63-dimensional representation is usually 
carried by a traceless  object with one up and one down index whose ranges
are one to eight; the count being $8.8-1=63$. However, in the formulation
we have in this paper it is instead  carried by
$J^I{}_J$ and
$S^{I_1\ldots I_4}$; the count being ${8.7\over 2}+ {8.7.6.5\over
2.4!}=63$. 
\par
In previous papers we have labeled representations of SL(n)
by giving the Dynkin diagram with the $n-1$ dots in a horizontal row
labeled from
$1$ to $n-1$ from left to right and then taking  the fundamental
representation associated with node $1$ to be the $\bar {n}$
representation and carried by the tensor $T_i, \ i=1,2,\ldots ,n$ while
the fundamental representation associated with node $n-1$ to be the $ {n}$
representation and carried by the tensor $T^i, \ i=1,2,\ldots ,n$. The
fundamental representation associated with node $n-2$ is then the $n(n-1)$
representation carried by the tensor $T^{i_1i_2}= T^{[ i_1i_2 ] }$, etc.
The use of over bars  also applies to the SL(n) Dynkin diagram  when
embedded into the 
$E_{n+1}$ Dynkin diagram, for example the fundamental representation
associated with node one of the $E_7$ Dynkin diagram is the $\bar {27}$.
However, given the unfamiliar way the SU(8) representations occur here we
take,  by definition,  
$T_{+}^{a}{}^{I_{1}I_{2}}$ to be the 28-dimensional representation of
SU(8). 
\par
The coset generators formed from the level two and minus two 
generators are given by 
$$
T^{a_{1}a_{2}I}{}_{J}= 
R^{a_{1}a_{2}I}{}_{J}+\tilde{R}_{a_{1}a_{2}}{}^{J}{}_I,\quad
T^{a_{1}a_{2}I_{1}\cdots I_{4}}= R^{a_{1}a_{2}I_{1}\cdots I_{4}}-\star
\tilde{R}_{a_{1}a_{2}}{}^{I_{1}\cdots I_{4}}, \quad
\hat T^{ab}= \hat K^{ab}+\tilde{\hat K}_{ab}
\eqno(2.4.7)$$
We note that $T^{a_{1}a_{2}K}{}_{K}=0$, but otherwise this object has no
particular symmetry. Similarly, $T^{a_{1}a_{2}I_{1}\cdots I_{4}}$ is
neither self-dual or anti-self dual. 
Their commutators with the SO(8) generators is as one expects
and those with the remaining SU(8) generators are given by 
$$
[S^{I_1\ldots I_4}, T_S^{a_1a_2} {}^J{}_K]=Ê
-2\delta_K^{[I_1 |} T_+^{a_1a_2}{}^{J|I_2I_3I_4]}
-2\delta_J^{[I_1|} T_+^{a_1a_2}{}^{K|I_2I_3I_4]}
+{1\over 2} \delta_K^J T_+^{a_1a_2}{}^{I_1I_2I_3I_4]}
$$
$$
[S^{I_1\ldots I_4}, T_A^{a_1a_2} {}^J{}_K]=Ê
-2\delta_K^{[I_1|} T_-^{a_1a_2}{}^{J|I_2I_3I_4]}
+2\delta_J^{[I_1|}  T_-^{a_1a_2}{}^{K|I_2I_3I_4]}
$$
$$
[S^{I_1\ldots I_4}, T_+^{a_1a_2} {}^{J_1\ldots J_4}]=Ê
{4\over 3}  \delta^{[J_1J_2J_3 |}_{[I_1I_2I_3}
T_S^{a_1a_2}{}^{|J_4]}{}_{I_4]} +{1\over 18}  \epsilon ^{I_1\ldots I_4
L[J_1J_2J_3 |} T_S^{a_1a_2}{}^{|J_4]}{}_{L}
$$

$$
[S^{I_1\ldots I_4}, T_-^{a_1a_2} {}^{J_1\ldots J_4}]=Ê
{4\over 3}  \delta^{[J_1J_2J_3 |}_{[I_1I_2I_3}
T_A^{a_1a_2}{}^{|J_4]}{}_{I_4]} +{1\over 18}  \epsilon ^{I_1\ldots I_4
L[J_1J_2J_3 |} T_A^{a_1a_2}{}^{|J_4]}{}_{L}
\eqno(2.4.8)$$
where 
$$
T^{ab}_S{}^{I}{}_{J} = {1\over 2} (T^{ab}{}^{I}{}_{J}+
T^{ab}{}^{J}{}_{I}) , \quad 
T^{ab}_A{}^{I}{}_{J} = {1\over 2} (T^{ab}{}^{I}{}_{J}-
T^{ab}{}^{J}{}_{I}), 
$$
$$
T^{ab}_\pm {}^{I_1I_2I_3I_4}= T^{ab} {}^{I_1I_2I_3I_4}\pm \star
T^{ab} {}^{I_1I_2I_3I_4}
\eqno(2.4.9)$$ 
We can interpret these equations as meaning thatÊ
$T_+^{a_1a_2} {}^{J_1\ldots J_4}$ and  $T_S^{a_1a_2} {}^J{}_K$ form the
$35+35=70$-dimensional representations of SU(8) whileÊ
$T_-^{a_1a_2} {}^{J_1\ldots J_4}$ and  $ T_A^{a_1a_2} {}^J{}_K$ form
the $35+28=63$-dimensional representations of SU(8).ÊWe recall that these
generators belong to the 133-dimensional representation of $E_7$. 
\par
The Cartan involution invariant subalgebra $I_c(E_{11})$ can be
constructed from the multiple commutators of  the generators of SU(8) and
the generators  $ S^{a}_{\pm}{}^{I_1I_2}$ of equation (2.3.10). As such to
know the commutators of the coset generators with all of those of the
Cartan involution subalgebra it suffices to find the commutators of the
coset generators with the generators $J^I{}_K, S^{I_1\ldots I_4}$ of SU(8)
and
$S^{a}_{\pm}{}^{I_1I_2}$. The former were given above and we now give the
commutators of the latter; the result with those of level zero of equation
(2.4.1) are given by 
$$
[ S^{aI_1I_2}_{\pm}, T^J{}_K]= -2\delta^{[I_1}_J T^a_\mp {}^{|K| I_2]}-
2\delta^{[I_1}_K T^a_\mp {}^{|J| I_2]} ,\quad
[ S^{aI_1I_2}_{\pm},
T^{b_1b_2} ]= -2 \delta ^{a(b_1} T^{b_2)}_\pm {}^{I_1I_2},
$$
$$
[ S^{aI_1I_2}_{\pm}, T^{J_1\ldots J_4} ]= \mp i\delta
^{[J_1J_2|}_{I_1I_2} T^a_\mp {}^{|J_3J_4]} \mp {i\over 4!}
\epsilon^{J_1\ldots J_4 I_1I_2 K_1K_2} T^a_\mp {}_{K_1K_2}
\eqno(2.4.10)$$
with the coset generators  of equation (2.4.4)  by 
$$
[ S^{aI_1I_2}_{\pm}, T^b_\pm {}^{J_1J_2} ]= 12 T^{ab}_+ {}_{I_1I_2J_1J_2}
+8 \delta ^a_b \delta ^{[I_1} _{[J_1} T^{I_2]}{}_{J_2]} \pm 24i  \delta
^a_b T^{I_1I_2J_1J_2}\mp 8i \delta ^{[I_1} _{[J_1}
T^{ab}_S{}^{I_2]}{}_{J_2]} 
$$
$$
[ S^{aI_1I_2}_{\pm}, T^b_\mp {}^{J_1J_2} ]= 12 T^{ab}_-{}^{I_1I_2J_1J_2}
+\delta ^{I_1I_2} _{J_1J_2}(4T^{ab}-2\sum_c T^c{}_c)
\pm 8i \delta ^{[I_1} _{[J_1}
T^{ab}_A{}^{I_2]}{}_{J_2] }
\mp4i \delta ^{I_1I_2} _{J_1J_2}\hat T^{ab}
\eqno(2.4.11)$$
While the commutators of the generators of equation (2.4.9) with
$S^{aI_1I_2}_{\pm}$, but keeping only those generators of levels plus and
minus one,  are given by 
$$
[ S^{aI_1I_2}_{\pm}, T^{b_1b_2}_S{}^J{}_K]= \mp 4i \delta _a^{[b_1}
\delta _{[I_1}^{(K|} T^{|b_2]} _\mp {}_{|J) |I_2 ]}\pm {i\over 2} 
\delta ^J_K \delta _a^{[b_1}
 T^{b_2]} _\mp {}_{I_1 I_2 }, 
$$
$$
[ S^{aI_1I_2}_{\pm}, T^{b_1b_2}_A{}^J{}_K]= \pm 4i \delta _a^{[b_1}
\delta _{[I_1}^{[J|} T^{|b_2]} _\pm {}_{|K] |I_2 ]} , 
\eqno(2.4.12)$$ 
$$
[ S^{aI_1I_2}_{\pm}, T^{b_1b_2}_+{}^{J_1\ldots J_4} ]= 2\delta
^{[b_1}_a\{ \delta ^{[J_1J_2 |}_{I_1I_2} T^{	|b_2]}_\mp {}^{|J_3J_4]} +
{1\over 4!}
\epsilon^{J_1\ldots J_4 I_1I_2 K_1K_2} T^{| b_2]}_\mp {}_{K_1K_2}\}
$$
$$
[ S^{aI_1I_2}_{\pm}, T^{b_1b_2}_-{}^{J_1\ldots J_4} ]= 2\delta
^{[b_1}_a\{ \delta ^{[J_1J_2 |}_{I_1I_2} T^{	|b_2]}_\pm {}^{|J_3J_4]} -
{1\over 4!}
\epsilon^{J_1\ldots J_4 I_1I_2 K_1K_2} T^{| b_2]}_\pm {}_{K_1K_2}\}
$$
$$
 [ S^{aI_1I_2}_{\pm}, \hat T^{b_1b_2} ]= \pm i \delta
^{a(b_1} T^{b_2)}_\pm {}^{I_1I_2}
\eqno(2.4.13)$$
\par
Finally we give the transformations of the $l_1$ representation under the 
Cartan involution invariant subalgebra. Rather than work with $Z^{I_1I_2}$
and $Z_{I_11I_2}$ we will work with the irreducible representations of
SU(8) which are given by 
$$
X_\pm{}^{I_1I_2}= Z^{I_1I_2}\pm i Z_{I_1I_2}
\eqno(2.4.14)$$
Using equation (2.2.22) we find that their  commutators with  the
generators of SU(8) are given by 
$$
[S^{I_1\ldots I_4}, X_\pm{}^{J_1J_2}]= \pm i  \delta _{[J_1J_2
|}^{[ I_1I_2}
 X_\pm{}^{I_3I_4]}]\mp {1\over 4!}\epsilon^{I_1\ldots I_4J_1J_2K_1K_2} 
X_\pm{}^{K_1K_2}
\eqno(2.4.15)$$
that is,  as the $28+\bar {28}$ of SU(8) should. 
\par
Using equations (2.2.23) and (2.2.24) we  also find that 
$$
[ S_\pm^a{}^{I_1I_2}, P_b]= \delta _b^a X_\pm{}^{I_1I_2},\quad 
[ S_\pm^a{}^{I_1I_2}, X_\pm^a{}^{J_1J_2}]=0,\quad
[ S_\pm^a{}^{I_1I_2}, X_\mp^a{}^{J_1J_2}]=-4  \delta _{J_1J_2}^{I_1I_2}P_a
\eqno(2.4.16)$$

\medskip
{\bf {3 The Cartan forms and  generalised vielbein}}
\medskip
We can finally  construct the   building blocks of the non-linear
realisation of
$E_{11}\otimes_sl_1$ appropriate to four dimensions, meaning the
semi-direct product algebra constructed from $E_{11}$ and its $l_1$
representations $l_1$. In this construction the commutators of
the generators of
$E_{11}$ with themselves are just those of $E_{11}$. The commutators of
generators of $E_{11}$ with those in the $l_1$ representations result
in generators in the $l_1$ representation and the Jacobi identities
then imply that the structure constants  are just the matrices of the
$l_1$ representation.  Clearly one can carry out this construction for
any group and one of its representations. Physicists are  very
familiar with semi-direct product algebras as the Poincar\'e algebra is
the semi-direct product of the translations and the Lorentz group. We take
the generators of the $l_1$ to commute, but more sophisticated
commutators  are possible. 
\par
We begin with a generic    group element
$g\in E_{11}\otimes_sl_1$ which can be written as 
$$
g=g_lg_E 
\eqno(3.1)$$
where 
$$
g_E= g_{-1}g_{-2}\ldots g_0\ldots g_2g_1
\eqno(3.2)$$
where $g_n$ contains level $n$  generators;  those with positive level 
are given by 
$$
g_0= e^{h_a{}^b K^a{}_b}e^{ \phi^I{}_J R^J{}_I}e^{\phi_{I_1\ldots
I_4}R^{I_1\ldots I_4}}\equiv g_h g_\phi,\quad 
g_1= e^{ A_{a}{}_{I_1I_2}R^{a}{}^{I_1I_2}+
A_{a}{}^{I_1I_2}R^{a}{}_{I_1I_2}},
$$
$$ 
g_2=e^{\hat h_{ab} \hat K^{ab}}  e^{ A_{a_1a_2}{}^I{}_J
R^{a_1a_2}{}^J{}_I}e^{A_{a_1a_2}{}_{I_1\ldots I_4}R^{a_1a_2}{}^{I_1\ldots
I_4}}\
\eqno(3.3)$$ 
In this and the next equation we have used the generators in their SL(8)
basis.  The group element formed from the generators of the
$l_1$ representations is given by  
$$
g_l= e^{x^a P_a} e^{x_{I_1I_2}Z^{I_1I_2}} 
e^{x^{I_1I_2}Z_{I_1I_2}}e^{\hat x_a Z^a} e^{x_{a}{}^I{}_J
Z^{a}{}^J{}_I}  e^{x_{a}{}_{I_1\ldots I_4} Z^{a}{}^{I_1\ldots I_4} }
\ldots =  e^{z^A L_A} 
\eqno(3.4)$$
where we have denoted the generalised coordinates by $z^A$ and the
generators of the $l_1$ representation by $l_A$. Thus the non-linear
realisation introduces a generalised space-time with the coordinates 
$$
x^a, x_{I_1I_2} , x^{I_1I_2}, \hat x_a, 
x_{a}{}^I{}_J, x_{a}{}_{I_1\ldots I_4}, \ldots 
\eqno(3.5)$$
The fields that occur in the group element $g_E$ are taken to depend on
the generalised space-time that is the coordinates of equation (3.5).
\par
The non-linear realisation is by definition  just a set of dynamical
equations, or Lagrangian,  that is invariant under the transformations 
$$
g\to g_0 g, \ \ \ g_0\in E_{11}\otimes _s l_1,\ \ {\rm as \  well \  as} \ \ \ g\to gh, \ \ \ h\in
I_c(E_{11})
\eqno(3.6)$$
The group element $g_0$ is a rigid transformation, that is a  constant, 
while $h$ is a local transformation, that is it depends on the
generalised space-time.  As the generators in $g_l$ form a representation
of $E_{11}$ the above transformations for $g_0\in E_{11}$ can be written
as 
$$
g_l\to g_0 g_lg_0^{-1}, g_E\to g_0 g_E\quad {\rm and }  \quad g_E\to g_E h
\eqno(3.7)$$
As a consequence the coordinates are inert under the  local
transformations but transform under the rigid  transformations as 
$$
z^A L_A\to g_0 z^AL_A g_0^{-1} = z^\Pi D(g_0^{-1})_\Pi {}^A L_A
\eqno(3.8)$$
\par
Using the local transformation we may bring $g_E$ into the form 
$$
g_E= g_0\ldots g_2g_1
\eqno(3.9)$$ 
Thus the theory contains 
the graviton field $h_a{}^b$, associated with the generators $K^a{}_b$
of   $GL(4)$, the 70 scalars  $\phi^I{}_J, \phi_{I_1\ldots I_4}$, 
associated  with the generators $K^I{}_J$ and $R^{I_1\ldots I_4}$
respectively, as well as the gauge fields
$ A_{a}{}_{I_1I_2}$,
$A_{a}{}^{I_1I_2}$,
$A_{a_1a_2}{}^I{}_J$ and 
$A_{a_1a_2}{}_{I_1\ldots I_4}$, associated with  the level one and 
two generators,   and in addition at level two we have the field 
$ \hat h_{ab}$ corresponding to the generator $\hat K^{a b}$ which
is the dual field of gravity [10].  The parameterisation of the group
element differs from that used in some earlier works on
$E_{11}$, but this does not affect any physical results.  
\par
As explained in the introduction,  the $l_1$ representation contains
all the brane charges and as it also  leads to the generalised
space-time there is  a one to one relation between the brane charges and
the coordinates of the generalised space-time.   Furthermore, 
for every field in $E_{11}$  there is a corresponding  element in the
$l_1$ representation. As such for every field there is an associated
coordinate in the generalised space-time and an associated brane. 
For example, the metric $h_a{}^b$ 
corresponds to the space-time translations, that is the  charge
$P_a$, which is carried by the point particle, or pp-wave, and has
associated coordinate
$x^a$, the dual graviton
$\hat h_{ab}$ corresponds to the charge $\hat Z^a$,  which is carried by 
the Taub-NUT solution, and has associated coordinate $\hat x_a$, the gauge
fields
$A_{a}{}_{I_1I_2}$ and 
$A_{a}{}^{I_1I_2}$ corresponds to the brane charges $Z^{I_1I_2}$ and
$Z_{I_1I_2}$,  which are the sources for  corresponding brane
solutions,  and the associated 
coordinates are $x_{I_1I_2}$ and
$x^{I_1I_2}$.  
\par
The dynamics is usually constructed from the Cartan forms 
${\cal V}=g^{-1}d g$ as these  are obviously inert under the $E_{11}$
rigid transformations of equation (3.5) and only transform under the local
transformations as 
$$
{\cal V}\to h^{-1} {\cal V} h+ h^{-1} d h
\eqno(3.10)$$
Hence if we use the Cartan forms, the  problem of finding a set of field
equations which are invariant under equation (3.6)  reduces to
finding a set that is invariant under the local subalgebra 
$I_c(E_{11})$,  that is the local transformations also given in 
equation (3.6) and so equation (3.10). 
\par
The  Cartan forms  can be written as  
$${\cal V}= {\cal V}_E+{\cal V}_l
\eqno(3.11)$$
where 
$$
{\cal V}_E=g_E^{-1}dg_E \quad {\rm and }\quad {\cal V}_l= g_E^{-1}(g_l^{-1}dg_l) g_E
\eqno(3.12)$$
The first part ${\cal V}_E$ is just  the Cartan form for $E_{11}$ while
${\cal V}_l$ is a sum of generators in the $l_1$ representation.  
Both ${\cal V}_E$ and ${\cal V}_l$ are invariant under rigid
transformations and under local transformations they change as 
$$ 
{\cal V}_E\to h^{-1}{\cal V}_E h + h^{-1} d h\quad {\rm and }\quad 
{\cal V}_l\to h^{-1}{\cal V}_l h 
\eqno(3.13)$$
\par 
Let us  evaluate the $E_{11}$ part of the Cartan form 
$$
{\cal V}_E = dz^\Pi G_{\Pi, \bullet} R^\bullet= 
G _{a}{}^b K^a{}_b+ \Omega^I{}_J K^J{}_I + \Omega_{I_1\ldots I_4}
R^{I_1\ldots I_4}
$$
$$
+ G_{a}{}_{I_1I_2}R^{a}{}^{I_1I_2}+
G_{a}{}^{I_1I_2}R^{a}{}_{I_1I_2}
$$
$$ 
+\hat G_{ab} \hat K^{ab}+  G_{a_1a_2}{}^I{}_J
R^{a_1a_2}{}^J{}_I+ G_{a_1a_2}{}_{I_1\ldots I_4}R^{a_1a_2}{}^{I_1\ldots
I_4}+\ldots 
\eqno(3.14)$$
where  $\bullet $ denotes the indices on the generators of  $E_{11}$.
Explicitly one finds that  
$$
G _{a}{}^b=(e^{-1}d e)_a{}^b,\quad G_\phi = g_\phi^{-1}d g_\phi\equiv
\Omega^J{}_I K^I{}_J + \Omega_{I_1\ldots I_4} R^{I_1\ldots I_4}
$$
$$
G_{a}{}_{I_1I_2}=\tilde D A_{a}{}_{I_1I_2}, \quad 
G_{a}{}^{I_1I_2}=\tilde D A_{a}{}^{I_1I_2},
$$
$$
G_{a_1a_2}{}^I{}_J = \tilde D  A_{a_1a_2}{}^I{}_J
-2 A_{[a_1|}{}_{LJ}\tilde DA_{|a_2]}{}^{LI}
-2 A_{[a_1|}{}^{LI}\tilde DA_{|a_2]}{}_{LJ}
$$
$$
G_{a_1a_2}{}_{I_1\ldots I_4}= \tilde D A_{a_1a_2}{}_{I_1\ldots I_4}
+ 6 A_{[a_1|}{}_{I_1I_2}\tilde D A_{|a_2]}{}_{I_3I_4 ]}-{1\over 4}
\epsilon _{ I_1\ldots I_4 J_1\ldots J_4}  A_{[a_1|}{}^{J_1J_2}\tilde D
A_{|a_2]}{}^{J_3J_4}
$$
$$
\hat G_{ab}= \tilde D \hat h_{ab } - A_{(a_1|}{}_{I_1I_2}\tilde
D A_{|a_2)}{}^{I_2I_3}+A_{(a_1|}{}^{I_1I_2}\tilde D
A_{|a_2)}{}_{I_2I_3}
\eqno(3.15)$$
 where $e_\mu{}^a \equiv (e^h)_\mu{}^a$ and  
$$
\tilde D A_{a}{}_{I_1I_2}\equiv  d A_{a}{}_{I_1I_2}
+ (e^{-1}d e)_{a}{}^{b}A_{b}{}_{I_1I_2}
+2\Omega ^J{}_{[I_1|} A_{a}{}_{J|I_2]}
+\Omega_{I_1\ldots I_4} A_{a}{}^{I_3I_4},
$$
$$
\tilde D A_{a}{}^{I_1I_2}\equiv  d A_{a}{}_{I_1I_2}
+ (e^{-1}d e)_{a}{}^{b}A_{b}{}^{I_1I_2}
-2\Omega ^{[I_1|}{}_J A_{a}{}^{J|I_2]}
-\Omega _{I_1I_2J_1 J_2}  A_{a}{}^{J_1J_2}
\eqno(3.16)$$
 with analogous expressions for other quantities. 
\par
Let us now evaluate the part of the Cartan form in equation
(3.11) containing the generators of the $l_1$ representation; we may
write it as  
$$
{\cal V}_l= g^{-1} d g= dz^\Pi E_\Pi{}^A L_A
$$
$$= 
g_E^{-1}(dx^aP_a+ dx_{I_1I_2}Z^{I_1I_2}+   dx^{I_1I_2}Z_{I_1I_2}+ Z^a
d\hat x_a+  x_{a}{}^I{}_J Z^{a}{}^J{}_I+ dx_{a}{}_{I_1\ldots
I_4}Z^{a}{}^{I_1\ldots I_4}+\ldots  ) g_E
$$
$$
=
E^a P_a + E_{I_1I_2}Z^{I_1I_2}+   E^{I_1I_2}Z_{I_1I_2} +\ldots 
\eqno(3.17)$$ 
where $E^A = dz^\Pi E_\Pi{}^A$. 
Using equations (2.2.22-24) we find that 
  ${ E}_\Pi{}^A$, viewed as   a matrix,  is given at low orders by 
$$
{ E}= 
\left(\matrix {(det e)^{-{1\over 2}}e_\mu{}^a&-(det e)^{-{1\over 2}}
e_\mu{}^c A_{c}{}_{J_1J_2}& - (det e)^{-{1\over 2}}e_\mu{}^c
A_{c}{}^{J_1J_2}\cr  0&{\cal N}^{I_1I_2}{}_{J_1J_2}& {\cal
N}^{I_1I_2}{}^{J_1J_2}
 \cr 0&{\cal N}_{I_1I_2}{}_{J_1J_2}& {\cal
N}_{I_1I_2}{}^{J_1J_2}
\cr}\right)
\eqno(3.18)$$
The matrix ${\cal N}$ is the vielbein in the scalar sector, that
is 
$g_\phi^{-1}  (dx_{I_1I_2}Z^{I_1I_2}+   dx^{I_1I_2}Z_{I_1I_2})g_\phi
\equiv dx\cdot {\cal N} \cdot l $.  
This illustrates the fact that  the  non-linear realisation
leads to a generalised space-time with a  generalised tangent space,
which for the four dimensional theory  consists of the usual tangent
space  of four-dimensional space-time, a
$56$-dimensional tangent space and then higher level tangent spaces. The
tangent space can be  read off from the
$l_1$ representation in an obvious way. 
 The tangent space group is $I_c(E_{11})$.   
At lowest level in four dimensions the tangent space group is 
$SO(4)\otimes SU(8)$ and the tangent vectors transform, at lowest
level, in the $4$  representation of of SO(4) 
 and the 
$28+\bar {28}$ representations of SU(8). It will prove advantageous to
express the tangent space in terms of objects that  transform into
themselves, that is, identify precisely,  the $28$ and $ \bar {28}$ of
SU(8).  To this end we can rewrite ${\cal V}_l$ at lowest order as 
$$
E^aP_a+E_{I_1I_2}Z^{I_1I_2}+   E^{I_1I_2}Z_{I_1I_2}= 
E^aP_a+ E_+{}_{I_1I_2}X_+{}^{I_1I_2}+   E_-{}^{I_1I_2}X_-{}_{I_1I_2}
\eqno(3.19)$$
using the generators of equation (2.4.14). Comparing terms we find that 
$$ 
E_\pm{}_{I_1I_2}={1\over 2} ( E_{I_1I_2} \mp i E^{I_1I_2})
\eqno(3.20)$$
As we see, the non-linear realisation
$E_{11}\otimes_sl_1$ automatically encodes a generalised geometry
equipped with a generalised vielbein which will be given explicitly at low
levels shortly.  
\par
Our task is to find a set of dynamics which is  invariant under the rigid
and local transformations of equation (3.7) and with this in mind we now
consider in more detail the transformations of the two parts of the
Cartan form beginning with the $E_{11}$ part, ${\cal V}_E$.  As noted
above the Cartan forms only transform  under  local $I_c(E_{11})$
transformations. It is useful to introduce the operation 
$g^*=(I_c(g))^{-1}$ on the group. While $I_c$ is an automorphism, i.e. on two group elements 
$I_c(g_1g_2)=I_c(g_1)I_c(g_2)$, the action of $*$ reverses the order, that is 
$(g_1g_2)^* =(g_2)^* (g_1)^* $. The action of $*$ on  the algebra is given by $A^*=-I_c(A)$ and $(AB)^*=B^*A^*$. A group element belonging to   $I_c(E_{11})$ obeys 
$h^*=h^{-1}$ and  the two transformations of equation (3.7) imply that 
$g^*\to h^{-1}g^* (g_0)^*$.  
We write the Cartan forms ${\cal V}_E$ as 
$$
{\cal V}_E=P+Q, \ \ {\rm where }\ \ P={1\over 2}({\cal V}_E+{\cal
V}_E^*),\ Q={1\over 2}({\cal V}_E-{\cal
V}_E^*)
\eqno(3.21)$$
and  then  the transformations of equation (3.13) become   
$$
P\to h^{-1}Ph,\ \ Q\to h^{-1}Qh+  h^{-1} dh
\eqno(3.22)$$
Examining equation (3.14)  we find that 
$$
2P= G ^b{}_{a} T^a{}_b +  \Omega^J{}_I T^I{}_J +
\Omega_{I_1\ldots I_4} T^{I_1\ldots I_4}
+ G_{a}{}_{I_1I_2}T^{a}{}^{I_1I_2}+
G_{a}{}^{I_1I_2}T^{a}{}_{I_1I_2}
$$
$$ 
+\hat G_{ab} \hat T^{ab}+  G_{a_1a_2}{}^I{}_J
R^{a_1a_2}{}^J{}_I+ G_{a_1a_2}{}_{I_1\ldots I_4}T^{a_1a_2}{}^{I_1\ldots
I_4}+\ldots 
$$
$$
=  G^b{} _{a} T^a{}_b +  \Omega^J{}_I T^I{}_J +
\Omega_{I_1\ldots I_4} T^{I_1\ldots I_4}
+ G_{a +}{}_{I_1I_2}T^{a}_+{}^{I_1I_2}+
G_{a -}{}_{I_1I_2}T^{a}_- {}^{I_1I_2}
$$
$$ 
+\hat G_{ab} \hat T^{ab}+  G_{a_1a_2 S}{}^I{}_J
T_S^{a_1a_2}{}^J{}_I+G_{a_1a_2 A}{}^I{}_J
T_A^{a_1a_2}{}^J{}_I
$$
$$
+  G_{a_1a_2}{}_{+I_1\ldots
I_4}T^{a_1a_2}_+{}^{I_1\ldots I_4}+G_{a_1a_2}{}_{- I_1\ldots
I_4}T^{a_1a_2}_-{}^{I_1\ldots I_4}+\ldots
\eqno(3.23)$$
and 
$$
2Q= G _{a}{}^b J^a{}_b +  \Omega^J{}_I J^I{}_J +
\Omega_{I_1\ldots I_4} S^{I_1\ldots I_4}
+ G_{a}{}_{I_1I_2}S^{a}{}^{I_1I_2}+
G_{a}{}^{I_1I_2}S^{a}{}_{I_1I_2}
$$
$$ 
+\hat G_{ab} \hat S^{ab}+  G_{a_1a_2}{}^I{}_J
S^{a_1a_2}{}^J{}_I+ G_{a_1a_2}{}_{I_1\ldots I_4}S^{a_1a_2}{}^{I_1\ldots
I_4}+\ldots 
\eqno(3.24)$$
Where in the first line we have used the generators of equations (2.4.1),
(2.4.4)  (2.4.7) and in the second line the  generators of
equations (2.4.5) and (2.4.9) which transform as irreducible
representation under SU(8). The Cartan forms inherit the properties of
the generators from which they arise; for example
$G_{a_1a_2}{}_{+I_1\ldots I_4}$ and $G_{a_1a_2}{}_{-I_1\ldots I_4}$ are
self dualand anti-self dual.  We note that except for the level zero
generators  the connection
$Q$ contains the same objects as the covariant quantity $P$. 
\par
Taking $h=1-\Lambda_{a +I_1I_2}S^{a}_+{}^{I_1I_2}- 
\Lambda_{a -I_1I_2}S^{a}_-{}^{I_1I_2}$, the local 
transformations of $P$ of equation (3.23) implies, 
using the equations of section four   that 

$$
\delta G^J{}_K =\sum_{\pm} ( 4 \Lambda _{a\pm}{}_{LK} G_{a\pm}{}_{L}{}^J+ 
4 \Lambda _{a\pm}{}_{LJ} G_{a\pm}{}_{L}{}^K
-\delta _K^J  \Lambda _{a\pm}{}_{LM} G_{a\pm}{}_{L}{}^M ) ,
$$
$$
\delta G^{I_1I_2I_3I_4} = 12i\sum_\pm \pm(\Lambda _{a\pm}{}_{[I_1I_2|
|}G_{a\pm}{}_{|I_3I_4]}+{1 \over 4!}\epsilon_{I_1\ldots I_4 }
{}^{J_1\ldots J_4} \Lambda _{a\pm}{}_{J_1J_2}G_{a\pm}{}_{J_3J_4})
$$
$$
\delta G_{a_1a_2} 
= \sum_\pm (4 \Lambda _{(a_1|\pm}{}_{I_1I_2 }G_{|a_2)\mp}{}_{I_1I_2}-2
\delta _{a_1a_2} \Lambda _{b\pm}{}_{I_1I_2 }G_{b\mp}{}_{I_1I_2})
$$
$$
\delta G _{a\pm}{}_{I_1I_2}= \pm 2iÊ \Lambda _{a\mp}{}_{J_1J_2 }
G^{I_1I_2J_1J_2}+4 \Lambda _{a\mp}{}_{J[I_1}G_{I_2]}{}^J +4\Lambda
_{b\mp}{}_{J_1J_2} G^b{}_{a}{}_{+J_1J_2I_1I_2}
$$
$$
+4\Lambda _{b\pm}{}_{J_1J_2} G^b{}_{a}{}_{-J_1J_2I_1I_2}
\pm 4i \Lambda _{b\mp}{}_{J[I_1} G^b{}_a{}_S{}^J{}_{|I_2]}Ê
\mp 4i \Lambda _{b\pm}{}_{J[I_1} G^b{}_a{}_A{}_{|I_2]}{}^JÊ
$$
$$
+\Lambda _{b\pm}{}_{I_1I_2 }(\pm i \hat G^b{}_a -2G^b{}_a)
$$
$$
\delta G_{a_1a_2}{}_{-I_1\ldots I_4}=6\sum_\pm (
Ê\Lambda _{[a_1|\pm}{}_{[I_1I_2 |} G_{|a_2]|\mp}{}_{|I_3I_4] }
-{1\over 4!} \epsilon_{I_1\ldots I_4}{}^{J_1\ldots J_4}
\Lambda _{[a_1|\pm}{}_{J_1J_2 } G_{|a_2]|\mp}{}_{J_3J_4 } ), 
$$
$$
\delta G_{a_1a_2 }{}_{+I_1\ldots I_4}=6\sum_\pm (
Ê\Lambda _{[a_1|\pm}{}_{[I_1I_2 |} G_{|a_2]|\pm}{}_{|I_3I_4] }
+{1\over 4!} \epsilon_{I_1\ldots I_4}{}^{J_1\ldots J_4}
\Lambda _{[a_1|\pm}{}_{J_1J_2 } G_{|a_2]|\pm}{}_{J_3J_4  }
$$
$$
\delta G_{a_1a_2 S}{}^J{}_K = \sum_\pmÊ \mp8i \Lambda _{[a_1|\pm}{}_{L
(K|} G_{|a_2]\pm}{}_{L}{}^{|J)}
 \pm i{\delta_K^J} \Lambda _{[a_1|\pm}{}_{L
M} G_{|a_2]\pm}{}_{L}{}^{M} , \quad 
$$
$$
\delta G_{a_1a_2 A}{}^J{}_K = \sum_\pmÊ \pm 8i \Lambda _{[a_1|\pm}{}_{L
[K|} G_{|a_2]\mp}{}_{L}{}^{|J]}
$$
$$
\delta \hat G_{a_1a_2}= \sum_\pmÊ \mp 4i \Lambda _{(a_1|\pm}{}_{J_1J_2}Ê
G_{|a_2)}{}_{\mp J_1J_2}
\eqno(3.25)$$
\par
Let us now turn our attention to the transformation of the
part of the Cartan form in the direction of the $l_1$ representation,
that is
${\cal V}_l$. At lowest level,  the transformation of equation (3.13)
implies, using equation (3.17) that 
$$
E_\Pi{}^{A\prime}= E_\Pi{}^B D(h)_B{}^A,\quad {\rm and \ for\  the \
inverse }\quad  (E^{-1})_A{}^{\Pi\prime}= D(h^{-1})_A{}^B (E^{-1})_B{}^\Pi
\eqno(3.26)$$
if we define $h^{-1}L_A h= D(h)_A{}^B L_B$.   
 At lowest levels this implies the local transforms 
$$
\delta E_\Pi{}^a= -4E_{\Pi}{}^-{}_{I_1I_2} \Lambda_{ a+}{}^{ I_1I_2}
-4 E_\Pi{}^+{}_{I_1I_2} \Lambda_{ a- }{}_{ I_1I_2}, \quad
\delta E_\Pi^\pm{}_{I_1I_2}=   \Lambda_{ a\pm}{}_{ I_1I_2}
E_\Pi{}^b,\ldots 
$$
$$
\delta (E^{-1})_a{}^\Pi=
-\Lambda_{a+}{}_{I_1I_2}(E^{-1}_+){}_{I_1I_2}{}^\Pi
-\Lambda_{a-}{}_{I_1I_2}(E^{-1}_-){}_{I_1I_2}{}^\Pi
$$
$$
 \delta (E^{-1}_\pm){}^{I_1I_2}{}^\Pi= 4\Lambda_{a\mp}
{}_{I_1I_2}(E^{-1})_b{}^\Pi
,\ldots 
\eqno(3.27)$$
\par
In the above we have written the Cartan 
forms as forms and were we to write them out explicitly we
would write
${\cal V}_E$ as 
$dz^\Pi G_{\Pi ,\bullet}R^\bullet$, where $\bullet$ denotes a generic
$E_{11}$ index, and   ${\cal V}_l$ as 
$dz^\Pi E_{\Pi}{}^A l_A$. Put another way we have suppressed their
world index $\Pi$. Even though the Cartan forms are invariant under the
rigid transformations,  
 $E_\Pi{}^A$ and $G_{\Pi ,  \bullet }$ are not as the transformation of
$z^\Pi$ of equation (3.8) implies a corresponding inverse transformation
acting on the $\Pi$ index of these two objects. Thus  $E_\Pi{}^A$
transforms under a local transformation on its $A$ index and by the
inverse of the coordinate transformation on its $\Pi$ index. As such we
can think of it as a generalised vielbein.  We can rewrite the Cartan
form of
$E_{11}\otimes_s l_1$ as 
$$
{\cal V}= g^{-1}d g= dz^\Pi E_\Pi{}^A( L_A+Ê G_{A,*} R^*)
\eqno(3.28)$$
where   $G_{A,\bullet}=(E^{-1})_A{}^\PiÊ G_{\Pi,\bullet}$. At low levels
$(E^{-1})_A{}^\Pi$ is the inverse of the matrix of equation (3.19).
ÊClearly $G_{A,\bullet }$ is inert under rigid transformations, but it 
transforms under local transformations as in equation (3.25) on its
$\bullet $ index  and as the inverse generalised vielbein on its $A$
index,  that is as in equation (3.27). The latter transformation can be
written as  
$$ 
\delta G_a{}_{, \bullet} =
-\Lambda_{a+}{}_{I_1I_2} G_{+I_1I_2}{}_{, \bullet}
-\Lambda_{a-}{}_{I_1I_2}G_{-I_1I_2}{}_{, \bullet}, \quad \delta
G_\pm{}_{I_1I_2}{}_{,\bullet} = 4\Lambda_{b\mp} {}_{I_1I_2}G_b {}_{,
\bullet}    ,\ldots 
\eqno(3.29)$$
Of course, the full local transformation  is
the sum of that given in equations (3.27) and (3.29). 
\par
The SU(8) variations of the Cartan forms is given on their  $E_{11}$ index
by taking $h=1-\Lambda ^{I_1\ldots I_4} R^{I_1\ldots I_4}$ in equation
(3.10) and  using equations (2.42),  (2.43),  (2.46) and  (2.48); the
result is Ê
$$
\delta G_{\diamond , }{}^I{}_J
= 6\Lambda_{K_1\ldots K_3 J} G_{\diamond , }{}^{K_1\ldots K_3 I}+Ê
6\Lambda_{K_1\ldots K_3 I} G_{\diamond , }{}^{K_1\ldots K_3 J}, \ Ê
\delta G_{\diamond , }{}^{J_1\ldots J_4}
= -16G_{\diamond , }{}^K{}_{[J_1|} \Lambda_{K|J_1\ldots J_4 ]}
$$
$$
\delta G_{\diamond , }{}_{\pm a}{}_{I_1I_2} 
=\pm 2i \Lambda _{J_1J_2I_1I_2}G_{\diamond , }{}_{\pm a}{}_{J_1J_2}
$$
$$
\delta G_{\diamond , }{}_{a_1a_2 S} {}^I{}_J=Ê-{8\over 3} \Lambda _{( I|
K_1K_2K_3} G_{\diamond, a_1a_2+}{}_{K_1K_2K_3 | J)}
$$
$$
\delta G_{\diamond , }{}_{a_1a_2 A} {}^I{}_J=-{8\over 3} \Lambda _{[ I|
K_1K_2K_3} G_{\diamond, a_1a_2-}{}_{K_1K_2K_3 | J]}
$$
$$
\delta G_{\diamond , }{}_{a_1a_2 +} {}^{I_1\ldots I_4}=-4\Lambda _{[
I_1I_2I_3 |K} G_{\diamond , }{}_{a_1a_2S} {}^K{}_{|I_4]}
$$
$$
\delta G_{\diamond , }{}_{a_1a_2 -} {}^{I_1\ldots I_4}=-4\Lambda _{[
I_1I_2I_3 |K} G_{\diamond , }{}_{a_1a_2 A} {}^K{}_{|I_4]}
\eqno(3.30)$$
While the Cartan  forms transform under SU(8) on their $l_1$, or $A$ index
as followsÊ
$$
\delta G_{\pm I_1I_2}{}_{, \bullet} = \mp 2i \Lambda _{I_1I_2 J_1J_2}Ê
G_{\pm J_1J_2} {}_{, \bullet}Ê
\eqno(3.31)$$

\medskip
{\bf {4 The equations of motion}}
\medskip
In this section we will construct the invariant equations of motion using
the variations found in the last section.  We found at the
end of section three that  the Cartan forms referred to the tangent space,
see equation (3.28),  are inert under the rigid $E_{11}$ transformations
and only transform under local $I_c(E_{11})$ transformations. 
Let us denote the Cartan forms in
${\cal V}_E$, when referred to tangent space,  by $G_{\diamond,\bullet}$ 
where $\bullet$ is a generic
$E_{11}$ index and $\diamond$ is a generic form index referred to the
tangent space using the generalised vielbein $E_A{}^\Pi$, 
in other words   $\diamond$ is the index used to label the $l_1$
representation.  The dynamics is by definition just a set of equations
which are  invariant under the local and rigid
transformations of the non-linear realisation.  Thus if  we  construct
the dynamics out of
$G_{\diamond ,\bullet}$ we need only worry about  the local
transformations.  Hence to find the dynamics  is just a problem in group
theory. However our knowledge of the properties of $I_c(E_{11})$ is
limited and so, for the time being, we  must carry this calculation out
level by level. We will  demand  that the equations of motion are
first order in derivatives and so first order in 
$G_{\diamond,\bullet}$.  This is a special feature of
$E_{11}$ reflecting the fact that $E_{11}$ is a duality symmetry
generalising electromagnetic duality. 
\par
The level zero Cartan involution invariant subalgebra is $SO(4)\times
SU(8)$ and so we can choose to classify the equations of motion by  
 representations of this algebra. The Cartan involution
invariant subalgebra is generated by  the level zero sector and the
generators $S_{a\pm I_1I_2}$ and so to check the invariance under the full
non-linear realisation  we need only check that the equations of motion
are inert under these  transformations. 
As such we will write down all terms, with arbitrary coefficients,  in the
chosen representation of $SO(4)\times SU(8)$,  up
to the level being studied,  and then vary them under the  transformations
of the Cartan involution invariant subalgebra; that is, the the SU(8)
transformations  given in equations (3.30) and (3.31) and the $S_{a\pm
I_1I_2}$ transformations given in equations (3.25) and (3.29). 
\par
Let us begin with the  equation of motion whose termsÊ belongs to the
6-dimensional  representation of SO(4), that is two antisymmetrised
indices,Ê and the 28 ($\bar {28}$)-dimensional representation of SU(8). 
While the representations of SO(4) are obvious the same is not always true
for those of SU(8), at least in the formalism we are using. However, in
constructing objects that transform as representations of SU(8)  we can be
guided by the well known action of the SO(8) subgroup.  
The most obvious  such terms are those Cartan forms whose $\bullet$ index,
that is
$E_{11}$ index,  carries, at least in part,  the
28 ($\bar {28})$-dimensional representation of SU(8) and whose
$\diamond $ index is just  the four dimensional
representation of GL(4), that is, the object $G_{[a_1, a_2]+}
{}_{I_1I_2} $ $(G_{a_1, a_2-} {}_{I_1I_2} )$. However, we can also
consider the Cartan forms whose $\bullet $ index contains the 
$70$-dimensional representation of
SU(8),  that is use the objects 
$G_{\diamond , }{}_{a_1a_2+}{}^{I_1\ldots I_4}$ and $G_{\diamond ,
}{}_{a_1a_2S}{}^{I}{}_{J}$,   and whose $\diamond$
index, that is $l_1$ index, belongs to the 28 ($\bar {28} $)
representation of SU(8), that is $G_{I_1I_2\pm , }{}_{\bullet}$.  We can
then form the 28 ($\bar {28}$) of SU(8) using the tensor product rules 
$28\times 70=\bar {28}+\ldots $ ($\bar {28}\times 70= {28}+\ldots $). 
Thus we
consider the sum of the two terms  
$$
Ê{ G}_{\pm J_1J_2 ,}{} _{a_1a_2 +} {}^{J_1J_2
I_1I_2 }, \quad { G}_{\pm [ I_1| K ,}{} _{a_1a_2 S} {}^{K}{}_{|I_2 ]}
\eqno(4.1)$$
Using the SU(8) variations of the Cartan forms of equations (3.30)
and (3.31) we find that the combination 
$$
{\Delta G}_{\pm a_1a_2 70} {}_{I_1I_2}\equiv 
Ê{ G}_{\pm J_1J_2 ,}{} _{a_1a_2 +} {}^{J_1J_2
I_1I_2 }\pm i { G}_{\pm [ I_1| K ,}{} _{a_1a_2 S} {}^{K}{}_{|I_2 ]}
\eqno(4.2)$$
transforms under SU(8) as 
$$
\delta ({\Delta G}_{\pm a_1a_2 70} {}_{I_1I_2})= 
\pm 2i \Lambda _{I_1I_2K_1K_2}{\Delta G}_{\pm a_1a_2 70} {}_{K_1K_2}
\eqno(4.3)$$
that is like the 28 ($\bar {28}$)-dimensional representation of SU(8)
and so like   $G_{\diamond , }{}_{a\pm }{}_{ I_1I_2}$. The use
of the subscript 70 reminds the reader of where this term originated and
it will be used to  distinguish this term  from a similar term that we
will also now construct. 
\par
We can also form the 28 ($\bar {28}$) representation by taking the
$\bullet $ index to be the 63-dimensional representation of SU(8) instead
of the 70-dimensional representation and using the tensor product rules 
$28\times 63=28+\ldots $ ($\bar {28}\times 63= \bar {28}+\ldots $). As
a result we consider the terms 
$$
Ê{ G}_{\pm J_1J_2 ,}{} _{a_1a_2 -} {}^{J_1J_2
I_1I_2 }, \quad { G}_{\pm [ I_1| K ,}{} _{a_1a_2 A} {}^{K}{}_{|I_2 ]}
\eqno(4.4)$$
Proceeding as before we find that the combination 
$$
{\Delta  G}_{\pm a_1a_2 63} {}_{I_1I_2}\equivÊ{ G}_{\mp J_1J_2 ,}{}
_{a_1a_2 -} {}^{J_1J_2 I_1I_2 }\mp i{ G}_{\mp [ I_1| K ,}{} _{a_1a_2
A} {}^{K}{}_{|I_2 ]}
\eqno(4.5)$$
transforms as 
$$
\delta ({\Delta G}_{\pm a_1a_2 63} {}_{I_1I_2})= 
\pm 2i \Lambda _{I_1I_2K_1K_2}{\Delta G}_{\pm a_1a_2 63} {}_{K_1K_2}
\eqno(4.6)$$
that is as the 28 ($\bar {28}$)-dimensional representation of SU(8). 
\par
Finally,  we can write down all possible terms in the equation of
motion that transforms as the 6-dimensional  representation of SO(4) and
the 28 ($\bar {28}$)-dimensional representation of SU(8); taking arbitrary
coefficients they are given by  
$$
G_{[a_1, a_2]\pm} {}_{I_1I_2} 
+i{e_{1\pm }\over 2} \epsilon _{a_1a_2} {}^{b_1b_2} G_{b_1, b_2\pm
}{}_{I_1I_2}Ê + e_{70\pm }{\Delta G}_{\pm}{} _{a_1a_2 70} {}_{I_1I_2}
+e_{63\pm }{\Delta G}_{\pm}{} _{a_1a_2 63} {}_{I_1I_2}
$$
$$
+i{e_{1\pm }\over 2} \epsilon _{a_1a_2} {}^{b_1b_2}
(e_{70\pm}^{\prime }{\Delta G}_{\pm}{} _{a_1a_2 70} {}_{I_1I_2}
+e_{63\pm}^{ \prime  }{\Delta G}_{\pm}{} _{a_1a_2 63} {}_{I_1I_2})
+\ldotsÊ=0
\eqno(4.7)$$
whereÊ $+\ldots $ mean terms at level greater that 
two in the fields andÊ derivativesÊ with respect to  the coordinates that
are greater than level zero.Ê
\par
Varying  equation (4.7) under a local transformation, but keeping only
terms that  contain the Cartan forms $G_{a_1, a_2\pm} {}_{I_1I_2}$, 
we find that the first term leads to a
term of the form 
$$
4G_{[a_1|,}{}^b{}_{|a_2]}{}_{+I_1I_2J_1J_2}\Lambda _{b\mp
J_1J_2}
\eqno(4.8)$$ 
as well as other terms. 
We can rewrite this term as 
$$
2G^{b}{}_{, }{}_{a_1a_2}{}_{+I_1I_2J_1J_2}\Lambda _{b\mp J_1J_2}
-6G_{[ b}{}_{, }{}_{a_1a_2]}{}_{+I_1I_2J_1J_2}\Lambda ^b{}_{\mp J_1J_2}
\eqno(4.9)$$ 
The first term can be canceled if we choose the constant in equation
(4.7) to be given by $e_{70\pm}= -{1\over 2}$. The second term is of the
form of a field strength and, as we will see, it is required to find the
original equation again after the variation.   We note that although
the local  variations of the Cartan forms do not lead to field strengths
for the gauge fields  there are allowed terms in the equation of motion
that   involve  derivatives with respect to  the extra coordinates which 
cancel the non-field strength terms.   Proceeding in the same way
for similar variations  one finds that 
$e_{63\pm}= -{1\over 2}$. Thus we find that, for these
coefficients, the variation of equation (4.7) is invariant if we discard 
variations that involve  other fields. Collecting these results we find
that equation (4.7) can be rewritten as 
$$
 {\cal G}_{[a_1, a_2]\pm} {}_{I_1I_2} 
\pm{i\over 2} \epsilon _{a_1a_2} {}^{b_1b_2} {\cal  G}_{b_1, b_2\pm
}{}_ {I_1I_2}=0
\eqno(4.10)$$
where 
$$
{\cal  G}_{[a_1, a_2]\pm}{}_ {I_1I_2}\equiv 
  G_{[a_1, a_2]\pm}{}_ {I_1I_2}
-{1\over 2} {\Delta G}_{\pm a_1a_2 70} {}_{I_1I_2}
-{1\over 2} {\Delta G}_{\pm a_1a_2 63} {}_{I_1I_2}
\eqno(4.11)$$
We recognise these as the correct equations of motion of the gauge fields 
once one acts with another derivative and takes the fields to depend only
on the usual coordinates of four-dimensional space-time. 
\par
When carrying out the variations in this section we consider  only
terms in the variations that have derivatives with respect to the usual
coordinates of space-time. As a result one finds new equations that
contain only terms with space-time derivatives. However, when we vary
these  equations the result is sensitive, by using equation (3.29), to 
terms in the original equation that contain 
derivatives with respect to the generalised coordinates, that is,  the
derivative 
$E_{\pm I_1I_2}{}^\Pi \partial_\Pi $. 
\par 
We now  take  all the other variations of equation (4.7) under
equations (3.25) and (3.29), except those involving the fields of gravity
and dual gravity.   We find that it leads to the equations 
$$
{ G}_{a, }{}^{I_1I_2J_1J_2} 
-\epsilon_{ab_1b_2b_3} G_{b_1, }{}_{b_2b_3+}{}^{I_1I_2J_1J_2}=0
\eqno(4.12)$$
$$
{G}_{a, }{}^{I}{}_J -\epsilon_{ab_1b_2b_3} G_{b_1,
}{}_{b_2b_3S}{}^{I}{}_J=0
\eqno(4.13)$$
In carrying out this calculation one must set to
zero the   coefficients of the parameters $\Lambda _{a+ I_1I_2}$ and 
 $\Lambda _{a- I_1I_2}$ as well as the independent  SO(1,3) tensor
structures  and in doing so one finds two copies of the above equations
that are only consistent if $e_{1\pm}^2=1$; in the above equation   we
have chosen $e_{1\pm}=\pm 1$. 
\par
Varying the  equations of motion for
the scalars (4.12) and (4.13) under the local $I_c(E_{11})$
transformations one finds the vector equation of motion of equation
(4.10).  However, as explained above,  it is in carrying out this step one
finds the contributions in equations (4.12) and (4.13) that contain the
derivatives with respect to the Lorentz scalar coordinates and the actual
equations now read 
$$
{\cal G}_{a, }{}^{I_1I_2J_1J_2} 
-\epsilon_{ab_1b_2b_3} G_{b_1, }{}_{b_2b_3+}{}^{I_1I_2J_1J_2}=0
\eqno(4.14)$$
$$
{\cal G}_{a, }{}^{I}{}_J -\epsilon_{ab_1b_2b_3} G_{b_1,
}{}_{b_2b_3S}{}^{I}{}_J=0
\eqno(4.15)$$
where 
$$
{\cal G}_{a, }{}^{I_1I_2J_1J_2} = {G}_{a, }{}^{I_1I_2J_1J_2} +6
\Delta{ G}_{a, }{}^{I_1I_2J_1J_2} 
,\quad {\cal G}_{a, }{}^{I}{}_J= {\cal G}_{a, }{}^{I}{}_J+6\Delta 
G_{a, }{}^{I}{}_J
\eqno(4.16)$$
and 
$$\Delta{ G}_{a, }{}^{I_1I_2J_1J_2} = + {i\over 2} (G_{-
[I_1I_2|,}{}_{ a+ |I_3I_4]} +{1\over 4!}  \epsilon_{I_1\ldots
I_4K_1\ldots K_4}  G_{- [K_1K_2|,}{}_{ a+ |K_3K_4] })
$$
$$
- {i\over 2} (G_{+
[I_1I_2|,}{}_{ a- |I_3I_4]} +{1\over 4!}  \epsilon_{I_1\ldots
I_4K_1\ldots K_4}  G_{+ [K_1K_2|,}{}_{ a- |K_3K_4] })
, 
$$
$$
\Delta G_{a, }{}^{I}{}_J= +{1\over 24} (G_{- IK, }{}_{a+ KJ}+G_{-
JK, }{}_{a+ KI}-{1\over 4} \delta^ I_JG_{- LK, }{}_{a+ KL}
$$
$$
+G_{+ IK, }{}_{a- KJ}+G_{+
JK, }{}_{a- KI}-{1\over 4} \delta^ I_JG_{+ LK, }{}_{a- KL})
\eqno(4.17)$$
One can verify that the combinations $\Delta{ G}_{a, }{}^{I_1I_2J_1J_2}$
and $\Delta G_{a, }{}^{I}{}_J$ transform as the 70-dimensional
representations of SU(8), that is, as ${ G}_{a, }{}^{I_1I_2I_3 I_4} $
and $G_{a, }{}^{I}{}_J$ do. 
\par
Taking another derivative acting on equations (4.14) and (4.15) we find
the equations of motion for the scalars of four dimensional maximal
supergravity provided we again take the fields to depend only on just the
usual coordinates of four dimensional space-time. 
\par
In varying equation (4.10) we also find the equations 
$$
G_{b_1, }{}_{b_2b_3-}{}^{I_1I_2I_3I_4}=0
\eqno(4.18)$$
$$
G_{b_1, }{}_{b_2b_3A}{}^{I}{}_J=0
\eqno(4.19)$$
These equations are expected as the fields in the non-linear realisation
which are dual to the scalars belong to the 133 of $E_7$,  however, there
are only 70 scalars as they belong to the non-linear realisation of $E_7$
with local subgroup SU(8), equivalently the coset $E_7$/SU(8). The
variations of these equations will be discussed later.  
\par
Thus one finds an infinite set of equations of motion that are invariant
under the symmetries of the non-linear realisation;  the lowest two
equations being the equations of motion for the gauge fields, equations
(4.10),   and scalars equations (4.14) and (4.15).  
 These latter equations  are equivalent to the equations
for the four dimensional maximal supergravity theory provided we take the
fields not to  depend on the Lorentz scalar coordinates.  We note
that these equations are uniquely determined by the symmetries once we
pick the Lorentz and SU(8) character of one of them. 
\par
We now consider equations whose Lorentz and SU(8) character do not occur
in the above infinite set. In particular we consider the equation that
is a Lorentz scalar but transforms under the 28 ($\bar {28}$) of SU(8).
Up to the level at which we are working the only possible terms that this
can contain are 
$$
G_{a, }{}^a{}_{\pm I_1I_2},\quad  G_{\pm J_1J_2,}{}^{J_1J_2I_1I_2} ,
\quad G_{\pm [I_1 |K,}{}^{K}{}_{|I_2]}
\eqno(4.20)$$
Varying under the local transformations of section three we find that 
this equation  will be invariant if it takes the form 
$$
G_{a, }{}^a{}_{\pm I_1I_2}\pm {i\over 2}  G_{\pm J_1J_2,}{}^{J_1J_2I_1I_2}
+{1\over 2}  G_{\pm [I_1 |K,}{}^{K}{}_{|I_2]}=0 
\eqno(4.21)$$
and we also impose the  additional  equations 
$$
G_{a,}{}^b{}_{a+I_1\ldots I_4} +{3\over 2} G_{-[I_1I_2|, }{}_{b+
|I_3I_4]} +{3\over 2} G_{+[I_1I_2|, }{}_{b -|I_3I_4]}=0 
\eqno(4.22)$$
$$
G_{a,}{}^b{}_{a-I_1\ldots I_4} +{3\over 2} G_{-[I_1I_2|, }{}_{b- |I_3I_4]}
+{3\over 2} G_{+[I_1I_2|, }{}_{b+ |I_3I_4]}=0 
\eqno(4.23)$$
$$
G_{a,}{}^b{}_{aS}{}^{J}{}_{K} -i G_{-L(K|b,}{}_{+L|J)} +i
G_{+L(K|b,}{}_{-L|J)}=0 
\eqno(4.24)$$
and 
$$
G_{a,}{}^b{}_{aA}{}^{J}{}_{K} +i G_{-L[K|b,}{}_{-L|J]} -i
G_{+L[K|b,}{}_{+L|J]}=0 
\eqno(4.25)$$
One can verify that the local variation of equations (4.22) to (4.25)
leads to equation (4.20).  Thus one finds another infinite tower of
invariant equations which are uniquely specified by the symmetries of the
non-linear realisation. 
\par
The equations (4.21-4.25)  can  be thought of as
gauge conditions, if one sets the dependence of the fields to be just
that of the coordinates of the usual four dimensional space-time. 
 Of course one does not have to actually adopt  these
latter equations  and one can just take the equations (4.10), (4.14), 
(4.15) and their higher level analogues. 
\par
We conclude this section with some incomplete results on the higher level
equations of motion. By varying the field equation for the gauge fields
we found equations (4.18) and (4.19). Varying the first equation under
the $I_c(E_{11})$ local variations we find that 
$$
8\sum_\pm (- \Lambda _{b\pm L[J|}G_{b, a \mp L|I]}+\Lambda _{b \pm
L[J|}G_{a, b \mp L|I]} )=0
\eqno(4.26)$$
The first term can be canceled by adding to the left-hand side of
equation (4.18) the term 
$$
2\sum_\pm G_{\mp L[J, a \mp L|I ]}
\eqno(4.27)$$
However, the second term can be canceled by adding the term 
$$
Q_{a ,}{}^{I}{}_{J}
\eqno(4.28)$$
Where $Q_{\diamond  ,}{}_{\bullet}$ are the Cartan forms belonging to
$I_c (E_{11})$. This has a local transformation which given by 
$$
\delta Q_{a ,}{}^{I}{}_{J}= -4\sum_\pm \Lambda _{a\pm
L[J|}Q_{a, b \mp L|I]} 
\eqno(4.29)$$
As we noted in the gauge in which we are working 
$Q_{a, b \mp L|I]}= G_{a, b \mp L|I]}$. However, $Q_{\diamond ,
}{}_{\bullet }$ does not transform homogeneously as it  has a $h^{-1}d h$
part. As such once we add terms of this type the equations of motion
only hold  modulo this inhomogeous term. Covariant equations can be found
by acting with a derivative in an appropriate way. The resulting equation
which replaces equation (4.18) is 
$$
\epsilon ^{ab_1b_2b_3}G_{b_1,b_2b_3 A}{}^{I}{}_{J}+2\sum_\pm G_{\mp L[J |,
a \mp}{{}^{ L|I ]}}+Q_{a ,}{}^{I}{}_{J}=0
\eqno(4.30)$$
\par
A similar analysis applies to equation (4.19) which is now replaced by 
the equation 
$$
\epsilon ^{a b_1b_2b_3}G_{b_1,b_2b_3 -}{}^{I_1\ldots I_4}-3i \sum_\pm
\mp G_{\mp [I_1I_2 |, a\pm |I_3I_4]} +{1\over 2} Q_{a
,}{}_{I_1\ldots I_4}=0
\eqno(4.31)$$
\par
The above steps are required in any non-linear realisation that is
constructed from a Kac-Moody algebra and involves scalars and has dual
fields. To illustrate this let us consider that the theory contains two
scalars that belong to the non-linear realisation of SL(2,R) which is
part of the larger Kac-Moody algebra. The theory  will also contain dual
fields which carry $D-2$ space-time indices,  if $D$ is the dimension of
space-time,  and belong to the adjoint representation of SL(2,R).
These lead to three field strengths which transform in the adjoint
representation of SL(2,R). However, only two of these are related to the
two scalars by a duality relation. This is possible as the Cartan forms
transform under the local symmetry which for Sl(2,R) is  SO(2). While two
of the Cartan forms are doublets the remainder is a singlet and this can
be set to zero, at least as far as the subalgebra SL(2,R) is concerned. 
The extra $D-2$ form field  arises as the $D-2$ forms must belong to a
multiplet of SL(2,R) while the scalars belong to the coset SL(2,R)/SO(2). 
However, the variation of this field under the other transformations of
the local subalgebra involves the other fields from the non-linear
realisation and these must be cancelled, hopefully in the way explained
above. One of the simplest contexts inwhich to think about this problem is
the IIB theory which has an obvious SL(2,R) subalgebra. 
\par
Varying the gauge field equation (4.10) but now keeping the gravity
and dual gravity fields we find that its real and imaginary parts are the
same and are given by 
$$
G_{[ a_1 |, }{}^{b}{}_{| a_2]}+{1\over 4} \epsilon_{a_1a_2}{}^{b_1b_2}
\hat G_{b_1 ,}{}^b{}_{b_2}=0
\eqno(4.32)$$
However to find the full equation one must vary this under the local
symmetry and then add the terms that contain derivatives with respect to
the generalised coordinates and the "$Q$" terms. The latter are the $Q_{a,
}{}^b{}_c$, that is the parts of the Cartan forms associated with the
Lorentz algebra. As a result the gravity equation will only hold up to
local Lorentz transformations which include a term which contains the
derivative of the Lorentz parameter. 
We note that the formulation of the correct gravity equation has some
features that are  similar to those for the scalar and this should 
increase the propect that the solution can to be found within the context
of the Kac-Moody algebra. 
\par
The resulting equation for gravity and those of equations (4.30) and
(4.31) are still being studied and the author expects to write a more
complete account in a subsequent publication. We
also hope to report on the significance of the second set of equations. 
\medskip
{\bf {5. Discussion}}
\medskip
One can view the above computation from a slightly different perspective. 
We have  considered  objects that are first order in the derivatives of
the generalised space-time. What we have shown, to the level to which we
are working,  is that the right-hand sides of equations (4.10), (4.14) 
and (4.15) vary into each other under the local symmetry $I_c(E_{11})$
and so transform covariantly.  Similarly, the left-hand side of
equations (4.21-4.25) vary into each other under  the local symmetry. 
Since the rigid symmetry is automatically encoded in the way we have
performed the computation it  follows that we have found two sets of 
expressions each of which transform covariantly under all the symmetries
of the non-linear realisation. Furthermore the two sets are uniquely
determined by the non-linear realisation,  that is,  the properties of 
the $E_{11}$ Kac-Moody algebra and its first fundamental representation
$l_1$. The only assumption we have made is that the objects we consider
are first order in the generalised space-time derivatives.  
\par
One is not forced to set either of these two sets of expressions to zero,
however, if one sets the first set to zero then one will find an
infinite number of equations the first two of which correctly describe the
equations of motion of the scalars and the gauge fields once we 
 consider the fields to depend only on the
usual coordinates of four dimensional space-time. 
\par
We can state the result in a more group theoretic manner. The Cartan
forms in the coset direction,  that is the $P$,  carry a representation
of $I_c(E_{11})$; the transformations acting on  the indices that are
inherited from  the adjoint representation of $E_{11}$ as well as  the
indices that arise from the $l_1$ representation. However, this
representation is not irreducible as there exists, at least at low levels,
an involution operator on the representation which
can be used to define the irreducible components.  This involution
includes the action of the  the epsilon symbol of the usual space-time
and it maps fields to their duals. Thus it  is a generalisation of our
usual notion of electromagnetic duality.  It would be good to understand 
the representation carried by the Cartan forms and the involution in
a more abstract way  as this could lead to a more efficient way of
computing the equations of motion rather than the order by order method
used in this paper. 
\par
We note that when we discard  coordinates from the the generalised
space-time except  those of the usual four-dimensional
space-time then  the equations are gauge invariant and  are unique so we
did not have  to adjust any constants in order to achieve this.  This is
in contrast to previous such computations in the early papers on $E_{11}$,
and for example in references [24,25], where one found the equations of
motion, or Lagrangian,  were only determined up to some constants. This
problem was addressed in the earliest $E_{11}$ papers by demanding that
the equations also be conformally invariant, an idea which was first used
in reference [19],  or by simply demanding gauge symmetry as in [24,25].
The difference with the calculation of this paper  is that one has
implemented the symmetries of the non-linear realisation at a higher
level and in particular the local symmetries which were often taken to be
just those at the very lowest level, that is, just the Lorentz group. 
\par
The results found in this paper are similar to  the calculation of
the $E_{11}\otimes_s l_1$ non-linear realisation in ten and eleven
dimensions given in references [29,30] and [31] respectively. Also in
these papers one considered quantities that were  first order in the 
derivatives with respect to  the generalised space-time and one found
only  two unique sets which transform covariantly into themselves under
the symmetries of the non-linear realisation.  Setting one of these sets
to zero leads to the equations of motion of the corresponding
supergravity.  The way the results in these papers were phrased were a
bit different,  but it is equivalent to the statement just made. 
\par
In the full non-linear realisation the field equations will depend on the
higher level fields which arise from the $E_{11}$ part of the non-linear
realisation and they will lead to effects which it would be interesting
to study. In particular we already know that the three form fields at
level four  will lead to the gauged supergravities in four dimensions
[12,36]. 
\par
It is striking, at least to this author, that the $E_{11}\otimes l_1$
non-linear realisation leads essentially uniquely to the correct equations
for the scalars and gauge fields and an equation, yet to be fully
formulated, for the gravity that has many of the correct features. Indeed
if the latter equation were to turn out to be  correct then the $E_{11}$
conjecture  would be proven. The present paper provides a good arena to
see if this is indeed the case. 
\par
The equations contain derivatives with respect to the higher level
coordinates  belonging to the generalised space-time. 
Just setting to zero the derivatives with respect to all the coordinates
except those of the usual four dimensional space-time is not a
satisfactory step. It is difficult to believe that the additional
coordinates, beyond those of the usual four dimensional space-time, are
not there for a reason. Indeed, as we have  already noted the higher
level coordinates do  play an important role in the formulation of
the gauged supergravities [13]. However, it remains to implement  a truly
satisfactory, physically motivated,  procedure  that carries out the
required radical reduction in the number of coordinates. We note that for
the
$E_{11}\otimes l_1$ non-linear realisation, at lowest level, in the
decomposition that leads to the IIA theory the reduction has been found
to occur by considering the first quantised theory [38]. While the full
twenty  dimensional generalised space-time occurs in the first quantised
theory its quantisation to find the field theory requires that half of the
coordinates  are eliminated. Thus the quantisation breaks  the manifest
SO(10,10) symmetry,  however, if one takes into account all possible ways
of choosing the ten dimensional  slice of  space-time that remains then
the theory should possess the full symmetry, albeit in a hidden way. It
would be interesting to extend these results to the full $E_{11}\otimes
l_1$ non-linear realisation. 
\par


\medskip
{\bf {Acknowledgment}}
\medskip 
I wish to thank Paul Cook and Michael Fleming for discussions and the
SFTC for support from grant number ST/J002798/1.


\medskip
{\bf {Appendix A The $E_{11} \otimes_sl_1$ algebra}}
\medskip 
This appendix is designed to equip the reader with the $E_{11}$ material
required to understand this paper. Rather than explain the theory behind
Kac-Moody algebras we will present the required results. We first  give
the
$E_{11}$ algebra in the decomposition appropriate to eleven dimensions,
that is, we decompose  the $E_{11}$ algebra into representations   of
$A_{10}$, or SL(11), representations [10,16]. This algebra is found by
deleting node eleven in the Dynkin Diagram of
$E_{11}$ given below
\medskip
$$
\matrix{
& & & & &&& &\otimes &11&&&
\cr & & & &&& & &| & && &
\cr
\bullet&-&\bullet&-&\ldots &- &\bullet&-&\bullet&-&\bullet&-&\bullet
\cr
1& &2& & & &7& &8& & 9&
&10\cr}
$$
\par
\centerline {Fig 1. The $E_{11}$ Dynkin diagram}
\medskip
The way one constructs this algebra from the definition of $E_{11}$ as a
Kac-Moody algebra in terms of representations of SL(11) is discussed, for
example,  in [17]. For the calculation in this paper one does not need to
understand all the subtleties of this construction and the parts of the
algebra that are needed are given below. The generators can be classified
according to a level which is associated with the decomposition
associated with the deletion of  node eleven.  At level zero we
have the algebra  GL(11) with the generators
$K^a{}_b,\ a,b =1,\ldots 11$ and at level one and minus one the rank
three generators $R^{abc}$ and 
$R_{abc}$ respectively.  The generators at level two and minus two are
$R^{a_1\ldots a_6}$
 and $R_{a_1\ldots a_6}$ respectively, while those at levels three and
minus three are $R^{a_1\ldots a_8,b}$
 and $R_{a_1\ldots a_8,b}$ respectively. The level is just the number of
upper minus lower indices divided by three. For a discussion giving the
more abstract definition of level which relates it to the deletion of
node eleven see for example references [17] or [38]. 
\par
The $E_{11}$ algebra at   levels zero and up  three is
given by [10,16]
$$
[K^a{}_b,K^c{}_d]=\delta _b^c K^a{}_d - \delta _d^a K^c{}_b, Ê
\eqno(A.1)$$
$$Ê [K^a{}_b, R^{c_1\ldots c_6}]=Ê
\delta _b^{c_1}R^{ac_2\ldots c_6}+\dots, \ Ê
Ê[K^a{}_b, R^{c_1\ldots c_3}]= \delta _b^{c_1}R^{a c_2 c_3}+\dots,
\eqno(A.2)$$
$$ [ K^a{}_b,Ê R^{c_1\ldots c_8, d} ]=Ê
(\delta ^{c_1}_b R^{a c_2\ldots c_8, d} +\cdots) + \delta _b^d
R^{c_1\ldots c_8, a} .
\eqno(A.3)$$
and 
$$[ R^{c_1\ldots c_3}, R^{c_4\ldots c_6}]= 2 R^{c_1\ldots c_6},\quad 
[R^{a_1\ldots a_6}, R^{b_1\ldots b_3}]
= 3Ê R^{a_1\ldots a_6 [b_1 b_2,b_3]},Ê
\eqno(A.4)$$
where $+\ldots $ means the appropriate anti-symmetrisation.Ê

The $E_{11}$ level zero  and negative level generators up to level minus
three obey the relationsÊ
$$
[K^a{}_b, R_{c_1\ldots c_3}]= -\delta ^a_{c_1}R_{b c_2
c_3}-\dots,\ [K^a{}_b, R_{c_1\ldots c_6}]=Ê -\delta ^a_{c_1}R_{bc_2\ldots
c_6}-\dots,
\eqno(A.5)$$
$$ [ K^a{}_b,Ê R_{c_1\ldots c_8, d} ]=Ê
-(\delta ^a_{c_1} R_{b c_2\ldots c_8, d} +\cdots) - \delta ^a_d
R_{c_1\ldots c_8, b} .
\eqno(A.6)$$
$$[ R_{c_1\ldots c_3}, R_{c_4\ldots c_6}]= 2 R_{c_1\ldots c_6},\quadÊ
[R_{a_1\ldots a_6}, R_{b_1\ldots b_3}]
= 3Ê R_{a_1\ldots a_6 [b_1 b_2,b_3]},Ê
\eqno(A.7)$$
Finally, the commutation relations between the positive and negative
generators Ê are given byÊ

$$[ R^{a_1\ldots a_3}, R_{b_1\ldots b_3}]= 18 \delta^{[a_1a_2}_{[b_1b_2}
K^{a_3]}{}_{b_3]}-2\delta^{a_1a_2 a_3}_{b_1b_2 b_3} D,\ Ê
[ R_{b_1\ldots b_3}, R^{a_1\ldots a_6}]= {5!\over 2}
\delta^{[a_1a_2a_3}_{b_1b_2b_3}R^{a_4a_5a_6]}
$$
$$
[ R^{a_1\ldots a_6}, R_{b_1\ldots b_6}]= -5!.3.3
\delta^{[a_1\ldots a_5}_{[b_1\ldots b_5}
K^{a_6]}{}_{b_6]}+5!\delta^{a_1\ldotsÊ a_6}_{b_1\ldotsÊ b_6} D ,\quadÊ
$$
$$
[ R_{a_1\ldots a_3}, R^{b_1\ldots b_8,c}]= 8.7.2
( \delta_{[a_1a_2 a_3}^{[b_1b_2b_3} R^{b_4\ldots b_8] c}-
Ê\delta_{[a_1a_2 a_3}^{[b_1b_2 |c|} R^{b_3\ldots b_8]} )
$$
$$
[ R_{a_1\ldots a_6}, R^{b_1\ldots b_8,c}]= {7! .2\over 3}
( \delta_{[a_1\ldotsÊ a_6}^{[b_1\dots b_6} R^{b_7 b_8] c}-
Ê\delta_{[a_1\ldotsÊ a_6}^{c[b_1\ldots b_5 } R^{b_6b_7 b_8]})
\eqno(A.8)$$
where $D=\sum_b K^b{}_b$, $\delta^{a_1a_2}_{b_1b_2}=
{1\over
2}(\delta^{a_1}_{b_1}\delta^{a_2}_{b_2}-
\delta^{a_2}_{b_1}\delta^{a_1}_{b_2})=
\delta^{[a_1}_{b_1}\delta^{a_2]}_{b_2}$ with similar formulae whenÊ
more indices are involved.Ê
\par
We also need the fundamental representation of $E_{11}$  associated with
node one, denoted by $l_1$. By definition this is the representation with
highest weight
$\Lambda_1$ which obeys 
$(\Lambda_1, \alpha_{  a})=\delta _{a,1}, \  a=1,2\ldots ,11$ 
where $\alpha_{ a}$ are the simple roots of $E_{11}$. In the
decomposition to Sl(11), corresponding to the deletion of node eleven, 
one finds that the $l_1$ representation  contains  the objects $P_a$, 
$Z^{ab}$ and
$Z^{a_1\ldots a_5} , a,b, a_1, \ldots =1,\ldots , 11$ corresponding to
levels zero, one and two respectively. We have taken the first object,
i.e.
$P_a$,  to have level zero by choice. Taking these to be generators
belong to  a semi-direct product algebra with those of $E_{11}$,   denoted
by $E_{11}\otimes _s l_1$,  their commutation relations with the level one
generators of $E_{11}$ are given by [16]
$$
[R^{a_1a_2a_3}, P_b]= 3 \delta^{[a_1}_b Z^{a_2a_3]}, \ Ê
[R^{a_1a_2a_3}, Z^{b_1b_2} ]= Z^{a_1a_2a_3 b_1b_2},\ Ê
$$
$$[R^{a_1a_2a_3}, Z^{b_1\ldots b_5} ]=Z^{b_1\ldots b_5[a_1a_2,a_3]}+
Z^{b_1\ldots b_5 a_1a_2 a_3}
\eqno(A.9)$$ 
These equations define the normalisation 
of the generators of the $l_1$ representation. The commutators of the
generators of the $l_1$ representation with those of GL(11) are given by 
$$
Ê[K^a{}_b, P_c]= -\delta _c^a P_b +{1\over
2}\delta _b^a P_c,\ Ê [K^a{}_b, Z^{c_1c_2} ]= 2\delta_b^{[c_1} Z^{|a|c_2]}
+{1\over 2}\delta _b^a Z^{c_1c_2},
$$
$$
[K^a{}_b, Z^{c_1\ldots c_5} ]= 5\delta_b^{[c_1} Z^{|a|c_2\ldots c_5]}
+{1\over 2}\delta _b^a Z^{c_1\ldots c_5}
\eqno(A.10)$$
The commutation relations with the level two generators of $E_{11}$ are
given by 
$$
[R^{a_1\dots a_6}, P_b]= -3 \delta^{[a_1}_b Z^{\ldots a_6]}, \ 
[R^{a_1\dots a_6}, Z^{b_1b_2} ]= Z^{b_1b_2[a_1\ldots a_5,a_6]},\ Ê
\eqno(A.11)$$
TheÊ commutators with the level $-1$ negative root generatorsÊare given
by 
$$
[R_{a_1a_2a_3}, P_b ]= 0,\ Ê
[R_{a_1a_2a_3}, Z^{b_1b_2} ]= 6\delta^{b_1b_2 }_{[a_1a_2} P_{a_3 ]},\ Ê
[R_{a_1a_2a_3}, Z^{b_1\ldots b_5} ]= {5!\over 2} \delta^{[ b_1b_2b_3
}_{a_1a_2a_3} Z^{b_4b_5]}
\eqno(A.12)$$

\medskip
{\bf {References}}
\medskip
\item{[1]} E. Cremmer and B. Julia, {\it The $N=8$  
supergravity theory. I. The Lagrangian.}, Phys.\ Lett.\ {\bf 80B} (1978)
48. 
\item{[2]} E. Cremmer and B. Julia, {\it The SO(8) Supergravity}  
Nucl Phys {\bf B159} (1979) 141. 
\item{[3]} N. Marcus and J. Schwarz, {\it Three-dimensional
supergravity theories}, Nucl. Phys. {\bf B228} (1983) 301. 
\item {[4]} H. Nicolai, The integrability of N=16 supergravity, 
Phys. Lett. 194B (1987) 402;   H.
Nicolai and N. Warner, {\it The Structure of $N=16$ Supergravity}, 
Commun. Math. Phys. {\bf 125} (1989) 369. 
\item {[5]}  B.\ Julia, {\it Group disintegrations}, in {\it  
Superspace and
Supergravity}, p. 331,  eds. S. W. Hawking  and M.  Ro\v{c}ek,  
Cambridge University
Press (1981); B. Julia, in {\it Vertex Operators in Mathematics  
and Physics},
Publications of the Mathematical Sciences Research Institute no 3,  
SpringerVerlag (1984).
\item{[6]} J, Schwarz and P. West,
{\it ``Symmetries and Transformation of Chiral
$N=2$ $D=10$ Supergravity''},
Phys. Lett. {\bf 126B} (1983) 301.
\item {[7]}   B. de Wit and H. Nicolai, {\it
D=11 supergravity   with local
$SU(8)$ invariance}, Nucl. Phys. {\bf B274} (1986) 363; H.  Nicolai,  
{\it Hidden
symmetries in D=11 supergravity},  Phys. Lett. {\bf 155B} (1985) 47;   
H.  Nicolai,
{\it D=11 supergravity with local $SO(16)$ invariance},  Phys. Lett.  
{\bf 187B}
(1987) 316;   S. Melosch and H. Nicolai, {\it New canonical  
variables for
$D=11$ supergravity}, Phys. Lett. {\bf B416} (1998) 91, {\tt
arXiv:hep-th/9709227};
\item {[8]}    K. Koepsell, H. Nicolai and H. Samtleben, {\it An
Exceptional    Geometry for $d=11$
Supergravity}, Class. Quant. Grav. {\bf 17} (2000) 3689, {\tt
arXiv:hep-th/0006034}. 
\item{[9]} P. West, {\it Hidden superconformal symmetries of  
M-theory}, {\bf JHEP 0008} (2000) 007, {\tt arXiv:hep-th/0005270}.
\item{[10]} P. West, {\it $E_{11}$ and M Theory}, Class. Quant.  
Grav.  {\bf 18}
(2001) 4443, {\tt arXiv:hep-th/ 0104081}; 
\item{[11]} I. Schnakenburg and  P. West, {\it Kac-Moody   
symmetries of
IIB supergravity}, Phys. Lett. {\bf B517} (2001) 421, {\tt  
arXiv:hep-th/0107181}.
\item{[12]}  F. ÊRiccioni and P. West, {\it
The $E_{11}$ origin of all maximal supergravities}, ÊJHEP {\bf 0707}
(2007) 063; ÊarXiv:0705.0752.
\item{[13]} ÊF. Riccioni and P. West, {\it E(11)-extended spacetime
and gauged supergravities},
JHEP {\bf 0802} (2008) 039, ÊarXiv:0712.1795
\item{[14]} ÊF. Riccioni and P. West,
Ê{\it Local E(11)}, JHEP {\bf 0904} (2009) 051, arXiv:hep-th/0902.4678.
\item{[15]} F. ÊRiccioni, ÊD. ÊSteele and P. West, {\it The E(11)
origin of all maximal supergravities - the hierarchy of field-strengths}
ÊÊJHEP {\bf 0909} (2009) 095, arXiv:0906.1177. 
\item{[16]} P. West, {\it $E_{11}$, SL(32) and Central Charges},
Phys. Lett. {\bf B 575} (2003) 333-342, {\tt hep-th/0307098}
\item {[17]} P. West, Introduction to Strings and Branes, Cambridge
University Press, June 2012. 
\item{[18]}  A. Kleinschmidt and P. West, {\it  Representations of G+++
and the role of space-time},  JHEP 0402 (2004) 033,  hep-th/0312247.
\item{[19]} V. Ogievetsky, Lett. {\it Infinite-dimensional algebra of
general  covariance group as the closure of the finite dimensional
algebras  of conformal and linear groups}, Nuovo. Cimento, 8 (1973) 988; 
 A. Borisov and V. Ogievetsky,  {\it Theory of dynamical affine and
conformal  symmetries as the theory of the gravitational field}, 
Teor. Mat. Fiz. 21 (1974) 32
\item{[20]} P. West,  {\it $E_{11}$ origin of Brane charges and U-duality
multiplets}, JHEP 0408 (2004) 052, hep-th/0406150. 
\item{[21]} P. West, {\it Brane dynamics, central charges and
$E_{11}$}, hep-th/0412336. 
\item{[22]} P. Cook and P. West, {\it Charge multiplets and masses
for E(11)}, ÊJHEP {\bf 11} (2008) 091, arXiv:0805.4451.
\item {[23]} P. West, {\it The IIA, IIB and eleven dimensional theories 
and their common
$E_{11}$ origin}, Nucl. Phys. B693 (2004) 76-102, hep-th/0402140. 
\item {[24]} C. Hillmann, {\it Generalized E(7(7)) coset dynamics and D=11
supergravity}, JHEP {\bf 0903}, 135 (2009), hep-th/0901.1581. 
\item{[25]} C. Hillmann, {\it E(7(7)) and d=11 supergravity }, PhD
thesis,  arXiv:0902.1509.
\item{[26]} D.  Berman, H. Godazgar, M. Perry and  P.  West, 
{\it Duality Invariant Actions and Generalised Geometry}, 
arXiv:1111.0459. 
\item{[27]}
D.~S.~Berman, M. Perry, {\it Generalized Geometry and M theory}, 
ÊÊJHEP {\bf 1106} (2011) 74
ÊÊ[arXiv:1008.1763 [hep-th]]; 
D.~S.Berman, H. Godazgar and M. Perry,
{\it SO(5,5) duality in M-theory and generalized geometry}, 
ÊÊPhys.\ Lett.\ Ê{\bf B700 } (2011) Ê65-67.
ÊÊ[arXiv:1103.5733 [hep-th]].
\item{[28]}  A. Tseytlin,  {\it Duality Symmetric Formulation
Of String World Sheet Dynamics}, Phys.Lett. {\bf B242} (1990) 163, {\it
Duality Symmetric Closed String Theory And Interacting Chiral Scalars},
Nucl.\ Phys.\ B {\bf 350} (1991) 395;  M. Duff, {\it Duality Rotations In
String Theory},
  Nucl.\ Phys.\  B {\bf 335} (1990) 610; M. Duff and J. Lu,
 {\it Duality rotations in
membrane theory},  Nucl. Phys. {\bf B347} (1990) 394. 
\item{[29]}  P. West, {\it E11, generalised space-time and IIA string
theory}, 
 Phys.Lett.B696 (2011) 403-409,   arXiv:1009.2624.
\item{[30]}   A. Rocen and P. West,  {\it E11, generalised space-time and
IIA string theory;  the R-R sector},  arXiv:1012.2744.
\item {[31]} P. West, {\it Generalised Geometry, eleven dimensions
and $E_{11}$}, arXiv:1111.1642.  
\item{[32]} O. Hohm, C. Hull and B. Zwiebach, {\it Generalised metric
formulation of double field theory},  hep-th/1006.4823; 
\item{[33]} N. Hitchin, Generalized Calabi-Yau manifolds, 
 Q. J. Math.  {\bf 54}  (2003), no. 3, 281,
math.DG/0209099;  {\it Brackets, form and
invariant functionals}, math.DG/0508618.
\item{[34]} M. Gualtieri, {\it Generalized complex geometry}, PhD Thesis
(2004), math.DG/0401221v1. 
\item{[35]} A. Coimbra, C. Strickland-Constable and  D.  Waldram, 
{\it Supergravity as Generalised Geometry I: Type II Theories}, 
arXiv:1107.1733; {\it $E_{d(d)} \times {R}^+$ Generalised Geometry,
Connections and M theory}, arXiv:1112.3989
\item {[36]} E. Bergshoeff, I. De Baetselier, T. Nutma, {\it E(11) and the
Embedding Tensor}, JHEP 0709 (2007) 047, arXiv:0705.1304. 
\item {[37]} P. West, {\it Very Extended $E_8$ and $A_8$ at low
levels, Gravity and Supergravity}, Class.Quant.Grav. {\bf 20} (2003)
2393, hep-th/0212291.
\item {[38]} P. West, {\it Generalised space-time and duality},
hep-th/1006.0893. 
\end